\documentclass[useAMS,usenatbib]{mn2e}
\usepackage{savesym}
\usepackage{txfonts}
\savesymbol{iint}
\savesymbol{iiint}
\savesymbol{iiiint}
\savesymbol{idotsint}
\usepackage{graphicx}
\usepackage{amssymb}
\usepackage[below]{placeins}
\usepackage{txfonts}
\usepackage{natbib}
\usepackage{amsmath}
\usepackage{float}
\restoresymbol{TXF}{iint}
\restoresymbol{TXF}{iiint}
\restoresymbol{TXF}{iiiint}
\restoresymbol{TXF}{idotsint}
\usepackage{multirow}
\usepackage{array}
\usepackage{xcolor}
\usepackage[percent]{overpic}
\bibpunct{(}{)}{;}{a}{}{,}
\def \arcmin{$^{\prime}$}

\def \xmm{{\emph{XMM-Newton}}}

\def \chandra{{\emph{Chandra}}}

\def\spose#1{\hbox to 0pt{#1\hss}}
\def\approxlt{\mathrel{\spose{\lower 3pt\hbox{$\sim$}}
        \raise 2.0pt\hbox{$<$}}}
\def\approxgt{\mathrel{\spose{\lower 3pt\hbox{$\sim$}}
        \raise 2.0pt\hbox{$>$}}}
\def\approxpropto{\mathrel{\spose{\lower 3pt\hbox{$\sim$}}
        \raise 2.0pt\hbox{$\propto$}}}
\mathchardef\twiddle="2218

\def\multleft#1{\hbox to size{\vbox {\halign {\lft{##}\cr #1}}\hfill}\par}
\def\multright#1{\hbox to size{\vbox {\halign {\rt{##}\cr #1}}\hfill}\par}

\def\today{\ifcase\month\or January\or February\or March\or April\or May\or
      June\or July\or August\or September\or October\or November\or December\fi
      \space\number\day, \number\year}
\def\<{\thinspace}

\def\ga{{\rm\thinspace gauss}}

\def\chandra{{\it Chandra}}

\def\xmm{{\it XMM-Newton}}

\def\arcmin {\hbox{$^{\prime}$}}

\makeatletter
\newcommand{\thickhline}{%
    \noalign {\ifnum 0=`}\fi \hrule height 1.2pt
    \futurelet \reserved@a \@xhline
}
\newcolumntype{"}{@{\hskip\tabcolsep\vrule width 1pt\hskip\tabcolsep}}
\makeatother

\newcommand{\ion}[2]{#1\,{\sc{#2}}}

\bibpunct{(}{)}{;}{a}{}{,}

\title[Resolving the nearest cold front in the sky]{Deep {\it Chandra} observation and numerical studies of the nearest cluster cold front in the sky 
}

\author[Werner et al.]{N. Werner$^{1,2}$, J.A. ZuHone$^{3}$, I. Zhuravleva$^{1,2}$, Y. Ichinohe$^{4,5}$, A. Simionescu$^{4}$, S.W. Allen$^{1,2,6}$, \newauthor M. Markevitch$^{7}$,  A.C. Fabian$^{8}$, U. Keshet$^{9}$, E. Roediger$^{10,11}$, M. Ruszkowski$^{12,13}$, J.S. Sanders$^{14}$ \\
$^1$Kavli Institute for Particle Astrophysics and Cosmology, Stanford University, 452 Lomita Mall, Stanford, CA 94305-4085, USA \\
$^2$Department of Physics, Stanford University, 382 Via Pueblo Mall, Stanford, CA 94305-4060, USA \\
$^3$MIT Kavli Institute for Astrophysics and Space Research, Massachusetts Institute of Technology, 77 Massachusetts Avenue, Cambridge, Massachusetts 02139, USA\\
$^4$Institute of Space and Astronautical Science (ISAS), JAXA, 3-1-1 Yoshinodai, Chuo-ku, Sagamihara, Kanagawa, 252-5210, Japan \\
$^5$Department of Physics, Graduate School of Science, University of Tokyo, 7-3-1 Hongo, Bunkyo, Tokyo 113-0033, Japan \\
$^6$SLAC National Accelerator Laboratory, 2575 Sand Hill Road, Menlo Park, CA 94025, USA \\
$^7$X-ray Astrophysics Laboratory, NASA Goddard Space Flight Center, Greenbelt, MD 20771, USA \\
$^8$Institute of Astronomy, Madingley Road, Cambridge CB3 0HA \\
$^9$Physics Department, Ben-Gurion University of the Negev, Be'er-Sheva 84105, Israel \\
$^{10}$Hamburger Sternwarte, Universit{\"a}t Hamburg, Gojensbergsweg 112, D-21029 Hamburg, Germany \\
$^{11}$Dublin Institute for Advanced Studies, Astronomy and Astrophysics Section, 31 Fitzwilliam Place, Dublin 2, Ireland \\
$^{12}$Department of Astronomy, University of Michigan, 500 Church Street, Ann Arbor, MI 48109, USA \\
$^{13}$Michigan Center for Theoretical Physics, 3444 Randall Lab, 450 Church St, Ann Arbor, MI 48109, USA \\
$^{14}$Max-Planck-Institut f{\" u}r extraterrestrische Physik, Giessenbachstrasse 1, 85748 Garching, Germany\\
}

\begin{document}
\maketitle

\begin{abstract}
We present the results of a very deep (500 ks) \chandra\ observation, along with tailored numerical simulations, of the nearest, best resolved cluster cold front in the sky, which lies 90 kpc (19\arcmin) to the northwest of M~87. 
The northern part of the front appears the sharpest, with a width smaller than 2.5~kpc (1.5 Coulomb mean free paths; at 99 per cent confidence). Everywhere along the front, the temperature discontinuity is narrower than 4--8 kpc and the metallicity gradient is narrower than 6 kpc, indicating that diffusion, conduction and mixing are suppressed across the interface.
Such transport processes can be naturally suppressed by magnetic fields aligned with the cold front. Interestingly, comparison to magnetohydrodynamic simulations indicates that in order to maintain the observed sharp density and temperature discontinuities, conduction must also be suppressed along the magnetic field lines.
However, the northwestern part of the cold front is observed to have a nonzero width. While other explanations are possible, the broadening is consistent with the presence of Kelvin-Helmholtz instabilities (KHI) on length scales of a few kpc. Based on comparison with simulations, the presence of KHI would imply that the effective viscosity of the intra-cluster medium is suppressed by more than an order of magnitude with respect to the isotropic Spitzer-like temperature dependent viscosity.
Underneath the cold front, we observe quasi-linear features that are $\sim10$ per cent brighter than the surrounding gas and are separated by $\sim15$ kpc from each other in projection. Comparison to tailored numerical simulations suggests that the observed phenomena may be due to the amplification of magnetic fields by gas sloshing in wide layers below the cold front, where the magnetic pressure reaches $\sim5$--10 per cent of the thermal pressure, reducing the gas density between the bright features.

\end{abstract}


\section{Introduction}

One of the first important results obtained with the  \chandra\ X-ray Observatory was the surprising discovery of {\it cold fronts} - remarkably sharp surface brightness discontinuities, with relatively dense and cool gas on the inner (bright) side and low density hot gas on the outer (faint) side. The first discovered cold fronts in Abell~2142 and Abell~3667 were interpreted as signs of mergers \citep{markevitch2000,vikhlinin2001}, where the sharp surface brightness discontinuity separates the hotter intra-cluster medium (ICM) from the low entropy core of a merging subcluster. 

However, as more clusters were observed with \chandra, cold fronts were also discovered in the centers of cool-core clusters (at $r=$10--400~kpc) with no signs of recent major mergers \citep{markevitch2001}. Indeed, such cold fronts are observed in most relatively relaxed cool-core clusters \citep{ghizzardi2010}, often with several, oppositely placed arcs curved around the central gas density peak, forming a spiral-like pattern \citep[see e.g.][]{markevitch2007,owers2009,simionescu2012,rossetti2013,ichinohe2015}. Considering projection effects and the weak density jumps, possibly all cool-core clusters may host some cold fronts. 

Cold fronts in cooling cores are believed to be due to the sloshing (or swirling) of the gas in the gravitational potential of the cluster. Numerical simulations show that sloshing can be induced easily by a minor merger where a subcluster falls in with a nonzero impact parameter \citep{tittley2005,ascasibar2006}. Due to this disturbance, the density peak of the main cluster swings on a spiral-like trajectory relative to its center of mass. As the central ICM is displaced from the corresponding dark matter peak, opposite and staggered cold fronts are created where cooler and denser parcels of gas from the centre come into contact with the hotter gas at larger radii. The large-scale coherent flows associated with sloshing and the corresponding cold fronts can persist for Gyrs. By redistributing the ICM, sloshing plays an important role in shaping the cluster cores and may even help prevent significant cooling flows \citep{zuhone2010,keshet2012}. 

\citet{markevitch2001} showed that sloshing cold fronts require significant acceleration of the gas beneath the contact discontinuity. This can be explained as centripetal acceleration of a tangential bulk flow just below the cold front \citep{keshet2010}, which causes a strong tangential shear along the discontinuity. This strong shear flow should promote the growth of hydrodynamic instabilities, which would broaden and eventually destroy the cold front on short sub-dynamical time scales. But cold fronts are remarkably sharp, both in terms of the density and the temperature jumps: the gas density discontinuity in e.g. the merging cluster Abell 3667 is at least several times narrower than the Coulomb mean free path \citep{vikhlinin2001,markevitch2007} and the observed temperature jumps require thermal conduction across cold fronts to be suppressed by a factor of order $\sim10^2$ compared to the collisional Spitzer or saturated values \citep{ettori2000,xiang2007}. A layer of magnetic field parallel to the discontinuity could stabilize the front, prevent the growth of instabilities, and suppress transport across the cold front \citep{vikhlinin2001b}. 

\citet{vikhlinin2001b} and \citet{vikhlinin2002} used the stability of the merger induced cold front in Abell~3667 to derive a lower limit on the magnetic field of $B>$7--16~$\mu$G at the discontinuity. 
Magnetic amplification in sloshing cold fronts is achieved through the shear due to the bulk flow of plasma beneath the front, which naturally sustains the strong magnetic fields parallel to cold fronts \citep{keshet2010,zuhone2011}. In the layers underneath cold fronts, fields with initial strengths of 0.1--1~$\mu$G may be amplified to tens of $\mu$G, and the magnetic pressures may be amplified to values reaching a significant fraction of the thermal pressure, leading to gas depleted bands \citep{zuhone2011}. 

\begin{figure}
\begin{minipage}{0.45\textwidth}
\vspace{-2.2cm}
\hspace{-0.7cm}
\includegraphics[width=1.22\textwidth,clip=t,angle=0.]{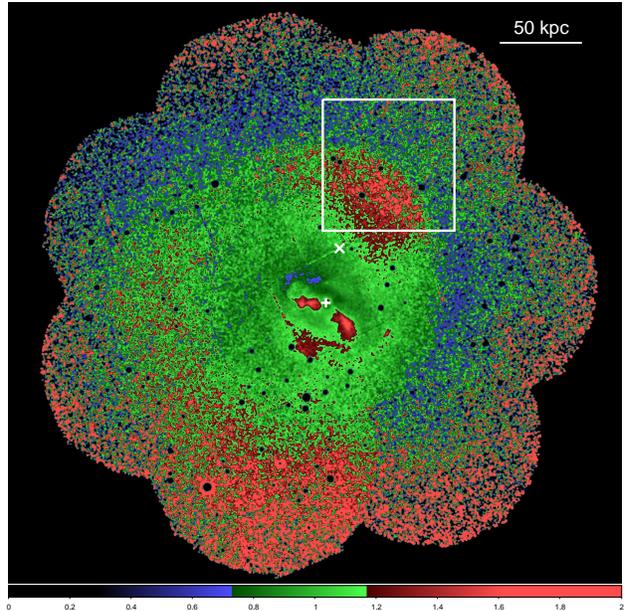}
\end{minipage}
\vspace{-2.2cm}
\caption{\xmm\ EPIC/MOS mosaic image of the central $r\sim150$~kpc region of the Virgo cluster. The exposure corrected image has been divided by the best-fitting radially symmetric beta model \citep[see][]{simionescu2010,rodiger2011}. The white square indicates our \chandra\ ACIS-I pointing shown in Fig.~2. The ``+'' sign indicates the center of M~87 at the bottom of the global gravitational potential of the Virgo Cluster. The ``X'' sign indicates the center of curvature (RA: 187.676, DEC: 12.5064) of the part of the cold front imaged by our  \chandra\ pointing.} 
\label{XMMim}
\end{figure}

One of the most interesting aspect of cold fronts is that they provide a relatively clean experimental setup. This allows us to address questions about the physics of the ICM in a more exact way than in cluster cores with complex astrophysical substructure. 
The closest example of a cluster cold front is the spectacular surface brightness discontinuity that lies 90 kpc (19\arcmin) to the north of M~87 in the Virgo Cluster and extends $140^{\circ}$ in azimuth \citep[see Fig.~\ref{XMMim},][]{simionescu2010}.  At the distance of 16.1~Mpc  \citep{tonry2001}, 1~arcsec corresponds to only 78~pc, making this the best resolved cluster cold front in the sky. The gas density jump across the discontinuity is $\rho_{\mathrm{in}}/\rho_{\mathrm{out}}\approx1.4$ and the gas in the inner, bright side is $\sim$40\% more abundant in Fe than the ICM outside the front \citep{simionescu2010}. \citet{simionescu2007} also detected a much weaker cold front $\sim 17$~kpc to the south-east of the core.  The opposite and staggered placement of these fronts, as well as the absence of a remnant sub-cluster, make gas sloshing the most viable explanation for the presence of these edges \citep{tittley2005,ascasibar2006}.

Using constrained hydrodynamical simulations, \citet{rodiger2011} showed that a sloshing scenario can reproduce the radii and the contrasts in X-ray brightness, projected temperature, and metallicity across the observed cold fronts in the Virgo Cluster. 
By comparing synthetic and real observations, they estimate that the original minor merger event that triggered the sloshing took place about 1.5~Gyr ago, when a sub-cluster of 2--$4\times10^{13}~M_{\odot}$ passed the Virgo core at 100 to 400 kpc distance nearly perpendicular to our line-of-sight. The resulting gas sloshing, with motions primarily in the plane of the sky, creates a contact discontinuity with a surface that is predominantly perpendicular to our line of sight, maximizing the contrast and minimizing projection effects. 
In subsequent work, \citet{rodiger2013} demonstrated the capability of deep \emph{Chandra} observations to reveal Kelvin-Helmholtz instabilities in the Virgo cold front by detecting multiple edges and shallower than expected surface brightness profiles across the discontinuity. Based on relatively short {\it XMM-Newton} observations, they also present tentative indications that the northeastern portion of the front is disturbed by instabilities. 

Here, we present a very deep (500 ks) {\it Chandra} observation of the sharpest, northwestern part of the Virgo cold front, designed to test the theoretical predictions in unprecedented detail.

\section{Observations and data analysis}

\begin{table}
\begin{center}
\caption{Summary of the \chandra\ observations.
Columns list the observation ID, observation date, 
and the exposure after cleaning.}\label{table:obs}
\begin{tabular}{ccc}
\hline\hline
Obs. ID & Obs. date   & Exposure (ks)\\
\hline 
15178& 2014 Feb. 17  & 47.1  \\
15179& 2014 Feb. 24  & 41.9  \\
15180& 2013 Aug. 1   & 140.5 \\
16585& 2014 Feb. 19  & 45.6  \\
16586& 2014 Feb. 20  & 49.8  \\
16587& 2014 Feb. 22  & 37.8  \\
16590& 2014 Feb. 27  & 38.1  \\
16591& 2014 Feb. 27  & 23.8  \\
16592& 2014 Mar. 1   & 36.1  \\
16593& 2014 Mar. 2   & 38.1  \\
\hline
\end{tabular}
\end{center}
\end{table}

\subsection{Chandra data}

The \chandra\ X-ray observatory observed the prominent cold front to the northwest of M~87 (see Fig.~\ref{XMMim}) between 2013 August and 2014 March using the Advanced CCD Imaging Spectrometer (ACIS-I).
The standard level-1 event lists were reprocessed using the {\texttt{CIAO}} (version~4.7, {\texttt{CALDB~4.6.5}}) software package, including the latest gain maps and calibration products. Bad pixels were removed and standard grade selections applied. The data were cleaned to remove periods of anomalously high background. The observations, along with the dates, identifiers, and net exposure times after cleaning (total net exposure of 500~ks), are listed in Table 1. Background images and spectra were extracted from the blank-sky fields available from the Chandra X-ray Center. These were cleaned in an identical way to the source observations, reprojected to the same coordinate system and normalized by the ratio of the observed to blank-sky count rates in the 9.5--12~keV band.

\label{analysis}
\begin{figure*}
\vspace{-1.5cm}
\begin{minipage}{0.45\textwidth}
\hspace{-1cm}
\includegraphics[width=1.17\textwidth,clip=t,angle=0.]{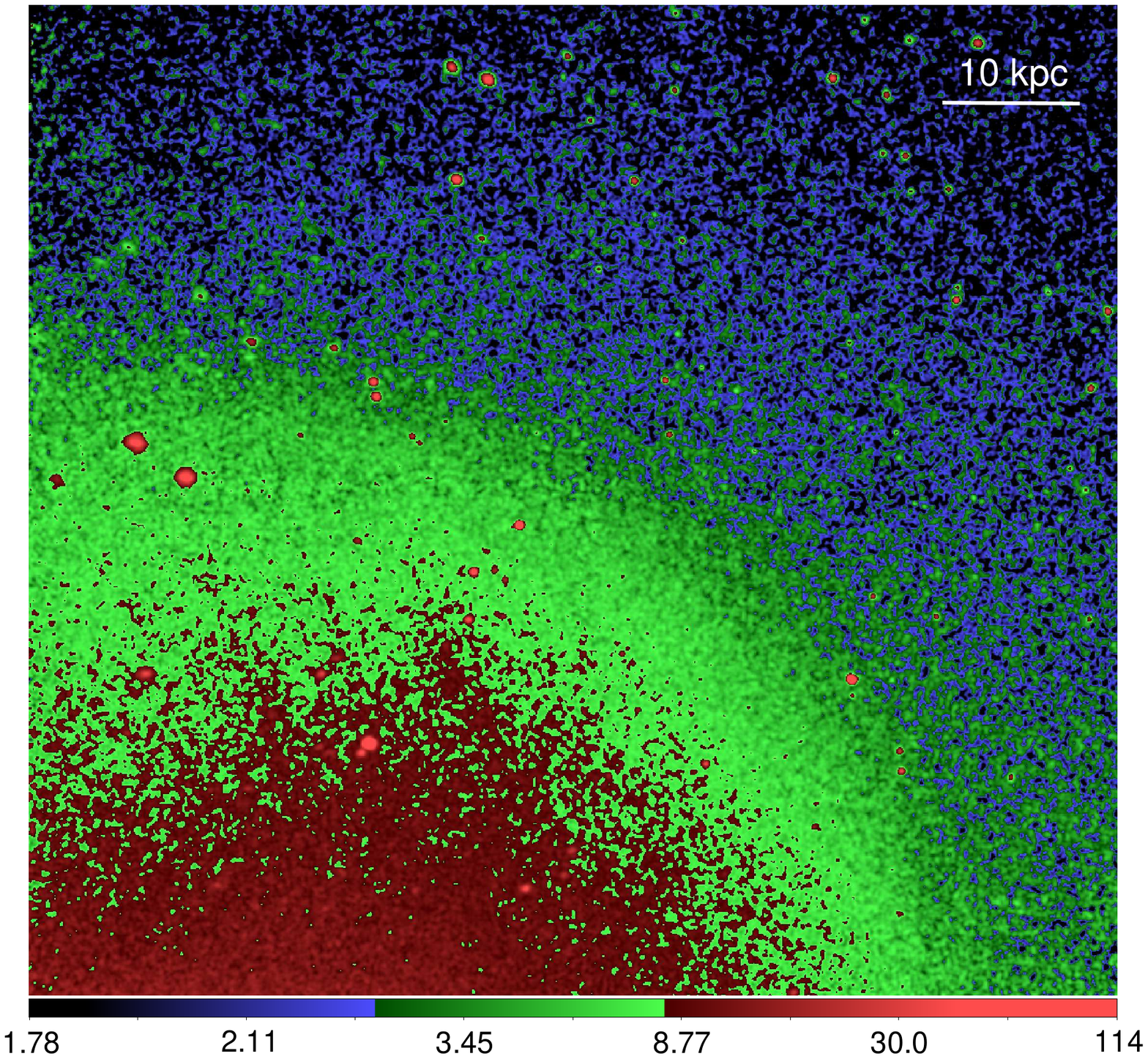}
\end{minipage}
\begin{minipage}{0.45\textwidth}
\includegraphics[width=1.17\textwidth,clip=t,angle=0.]{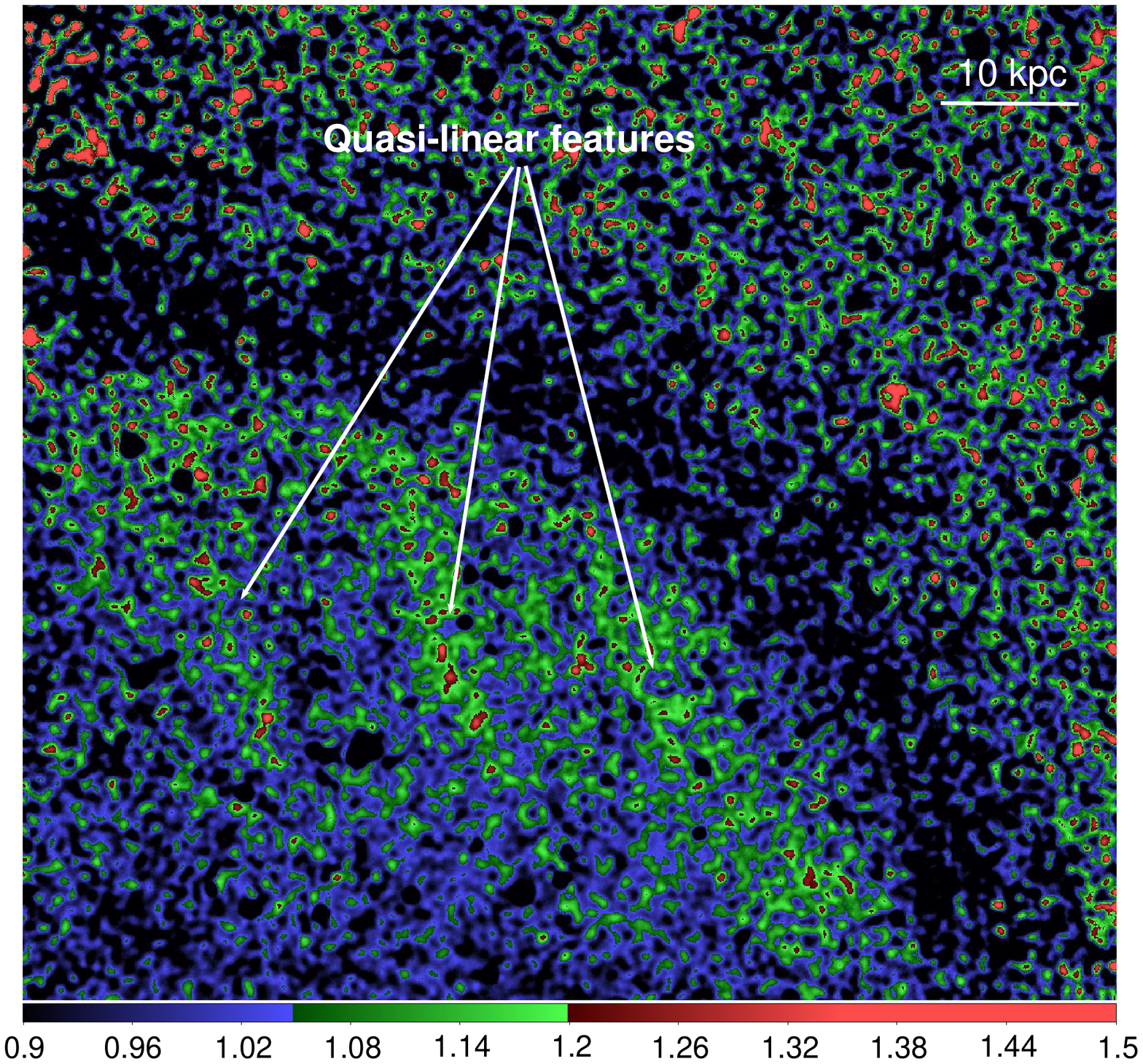}
\end{minipage}
\vspace{-1.5cm}
\caption{{\it Left panel:} \chandra\ image of the cold front in the 0.8--4.0~keV energy band. The image was smoothed with a Gaussian function with a 1.5 arcsec window. {\it Right panel:} The residual image obtained by dividing the image on the left with a beta-model, and patched on large scales with an 80~arcsec scale smoothing window \citep[see][and the text for details]{zhuravleva2015}, reveals three X-ray bright quasi-linear features separated from each other by $\sim15$~kpc in projection. The dark band outside the cold front is an image processing artifact due to the sharp surface brightness discontinuity. Point sources were removed from the residual image.  } 
\label{chandraim}
\end{figure*}

\subsubsection{Image analysis}

Background-subtracted images were created in nine narrow energy bands (0.8--1.0~keV, 1.0--1.2~keV, 1.2--1.5~keV, 1.5--2.0~keV, 2.0--2.5~keV, 2.5--2.75~keV, 2.75--3.0 keV, 3.0--3.5~keV, and 3.5--4.0~keV), spanning 0.8--4.0 keV. The nine narrow-band images were flat fielded with respect to the median energy for each image and then co-added to create the X-ray images shown in Fig.~\ref{chandraim}. In the 0.8--4.0~keV band the signal over background ratio is optimal and the emissivity is nearly independent of the gas temperature, maximizing the contrast at which we detect the contact discontinuity. Identification of point sources was performed using the {\texttt{CIAO}} task {\texttt{WAVDETECT}}. The point sources were excluded from the subsequent analysis. 

In order to emphasize structure in the image, we also produced a beta-model subtracted image, shown in the right panel of Fig.~\ref{chandraim}. In order to remove the underlying large-scale surface brightness gradient, we fitted a beta-model to an azimuthally averaged X-ray surface brightness profile extracted from the {\it XMM-Newton} mosaic image (see Sect. \ref{XMManalysis} and Fig.~\ref{XMMim}) and divided our {\it Chandra} image by this model, patched on large scales with an 80~arcsec smoothing window \citep[see][for details]{zhuravleva2015}.  Our patched model with which we divide our image is defined as $I_{\rm pm}=I_{\beta}S_{\sigma}[I_{\rm X}/I_{\beta}]$, where $I_{\beta}$ is the best fit beta-model to the azimuthally averaged X-ray surface brightness profile, $S_{\sigma}[.]$ is Gaussian smoothing with window size $\sigma$, and $I_{\rm X}$ is the X-ray surface brightness image.

\subsubsection{Spectral analysis}

Individual regions for the 2D spectral mapping were determined using the contour binning algorithm \citep{sanders2006b}, which groups neighbouring pixels of similar surface brightness until a desired signal-to-noise ratio threshold is met. We adopted a signal-to-noise ratio of 75 (approximately 5600 counts per region), which provides approximately 3 per cent precision for density, 4 per~cent precision for temperature, and 10--20 per cent precision for metallicity measurements, respectively. We modelled the spectra extracted from each spatial region with the {\texttt{SPEX}} package \citep{kaastra1996}. The spectral fitting was performed in the 0.6--7.5 keV band. The spectrum for each region was fitted with a model consisting of an absorbed single-phase plasma in collisional ionization equilibrium, with the spectral normalization (emission measure), temperature, and metallicity as free parameters. The deprojected spectra are obtained following the method described by \citet{churazov2003} and are fitted using {\texttt{XSPEC}} and the {\texttt{APEC}} plasma model \citep{smith2001,foster2012}. The line-of-sight absorption column density was fixed to the value, $N_{\rm H}=2\times10^{20}$~cm$^{-2}$, determined by the Leiden/Argentine/Bonn radio survey of \ion{H}{i} \citep{kalberla2005}. Metallicities are given with respect to the proto-Solar abundances by \citet{lodders2003}. 

\begin{figure*}
\vspace{0cm}
\includegraphics[width=\textwidth,clip=t,angle=0.]{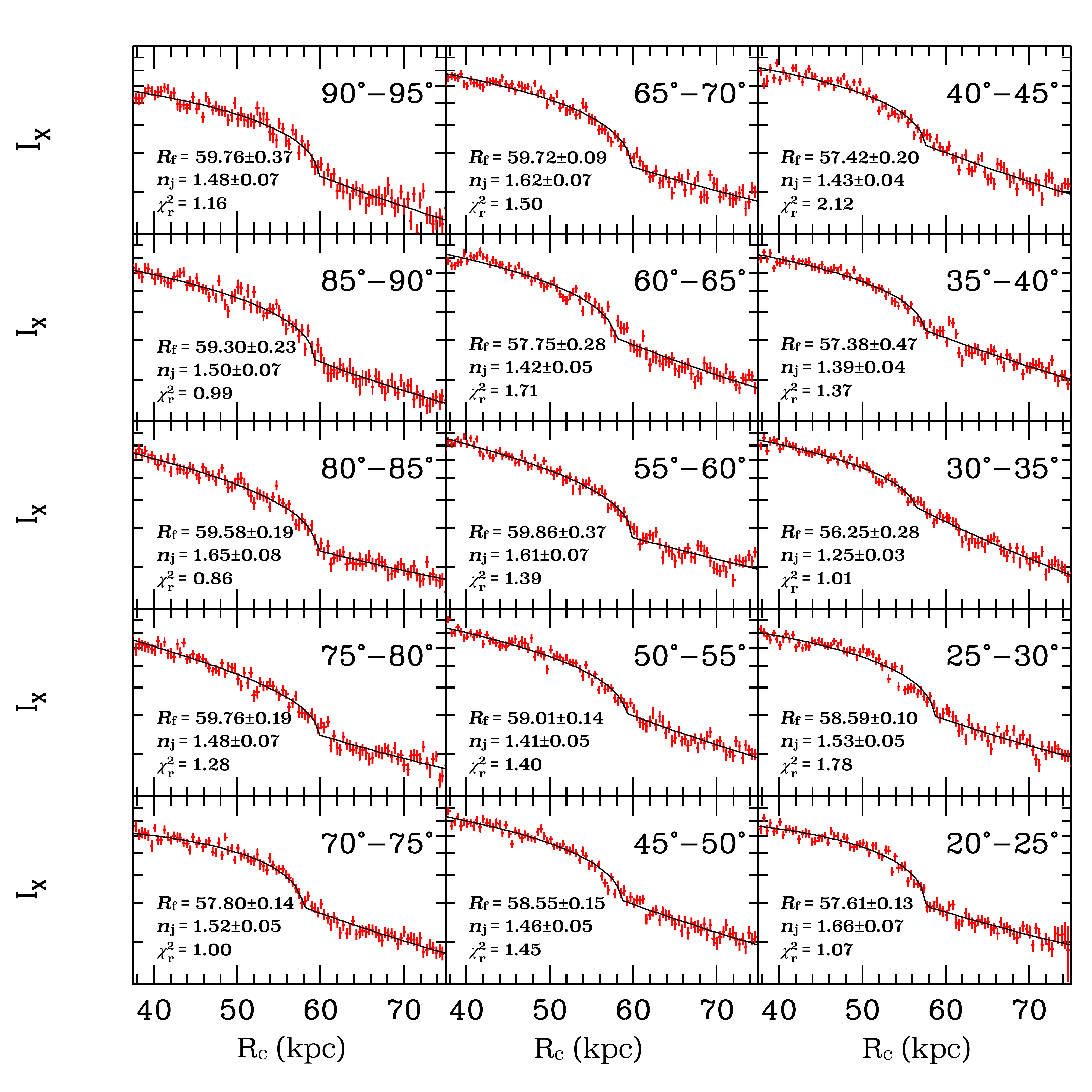}
\vspace{-0.5cm}
\caption{Surface brightness profiles across the cold front measured along 15 narrow, 5 degree wedges, each corresponding to $\sim5$~kpc along the front. The azimuth angles are measured counterclockwise from the west. Radii are measured from the center of curvature of the cold front (see Fig.~\ref{XMMim}). Over-plotted are the best fit broken power-law models for the ICM density profiles. For each wedge we indicate the best-fit radius of the discontinuity ($R_{\rm f}$, measured from the center of curvature of the cold front), density jump ($n_{\rm j}$), and the reduced $\chi^2$ for 92 degrees of freedom.
} 
\label{sbprofs}
\end{figure*}

\subsection{XMM-Newton data}
\label{XMManalysis}

A mosaic of six {\it XMM-Newton} pointings, each with an exposure time of $\sim20$~ks, was performed in June 2008. An additional pointing was obtained in November 2008. These observations cover an annulus around the deep (net exposure time of 120~ks) pointings on M~87. The full exposure corrected mosaic of {\it XMM-Newton} observations divided by the best fit radially symmetric beta-model is shown in Fig.~\ref{XMMim}. See \citet{simionescu2010} for the detailed discussion on the data and the analysis.

\section{Results}
\label{results}

The most prominent sloshing cold front in the Virgo Cluster extends at $r\sim 90$~kpc north of M~87 (see Fig.~\ref{XMMim}). Our \chandra\ observation (indicated by the white square in the figure) targets the part of the front with the largest surface brightness contrast. The deep {\it Chandra} image, on the left panel of Fig.~\ref{chandraim}, shows clearly this surface brightness discontinuity. 

The {\it Chandra} image also reveals unexpected, relatively bright quasi-linear features underneath the discontinuity. To our knowledge, this is the first time that such substructure has been identified with a cold front. Three of these quasi-linear features can be seen clearly on the residual image on the right panel of Fig.~\ref{chandraim}. They are $\sim10$~per cent brighter than the surrounding gas and are separated by $\sim15$~kpc  from each other in projection. These features are below the detection limit of the existing {\it XMM-Newton} observation. 

To study the detailed cold front interface, we extracted surface brightness profiles along 15 narrow, 5~degree wedges, each corresponding to $\sim5$~kpc along the front (see Fig.~\ref{sbprofs}). To optimize the sharpness of the surface brightness edge in our profiles, we aligned our annuli with the front interface, choosing the vertex of the wedges at the center of curvature of the highest-contrast part of the cold front (RA: $187^\circ.676$, DEC: $12^\circ.5064$). The data were binned into 4.92 arcsec radial bins. We fit the surface brightness profiles by projecting a spherically symmetric discontinuous, broken power-law density distribution \citep[see e.g.][]{owers2009}. The fitting was performed using the {\texttt{PROFFIT}} package \citep{eckert2011} modified by \citet{ogrean2015}. We fit the data in the radial range of 8--16 arcmin, corresponding to 38--75 kpc from the center of curvature. The free parameters of our model are the normalization, the inner and outer density slopes, and the amplitude and radius of the density jump. We verified that the best fit parameters are not sensitive to the utilized radial range. The surface brightness profiles over-plotted with their best fit models, along with the values for the best fit density jumps, radii of the discontinuity, and reduced $\chi^2$s (for 92 degrees of freedom) are shown in Fig.~\ref{sbprofs}.  

Our model provides the best fit to the northern part of the front, at azimuthal angles of 70--95 degrees, indicating that this is the sharpest part of the cold front. In particular, our model provides the best fit to the profile extracted at the azimuthal range of 80--85 degrees. In order to place an upper limit on the width of the front, we fit this profile with a model that includes Gaussian smoothing as an additional free parameter. We find that the 99 per cent upper limit on the smearing of this part of the cold front is 2.5 kpc, which is 1.5 Coulomb mean free paths (see Sect.~\ref{diffusedisc} for details). 
Given the quasi-spherical geometry along our line of sight, in this nearby system we observe a steep, few kpc wide surface brightness gradient at the front (in the more distant systems where the gradient cannot be resolved we only see a sharp surface brightness jump). Because of this resolved gradient, our upper limit on the width is significantly larger than {\it Chandra's} resolution.

On the other hand, at azimuths of 25--70 degrees, our model mostly provides a poorer fit to the data. Fig.~\ref{sbprofs} shows that around the brightness edge most profiles display deviations from the model, with shallow, smeared discontinuities. The parameters of the best fit model and the structure seen in the profiles also show azimuthal variation. In particular, the northwestern part of the front, at azimuths below 55 degrees, shows significant variation both in the value of the best fit density jump and radius.
For the sectors with the least sharp edges the fit significantly improves after including Gaussian smoothing in the model.  For example, for the profile extracted in the 65--70 degree wedge, Gaussian smoothing with a best fit width of 4.3 kpc improves the $\chi^2$ from 138 (for 92 degrees of freedom) to 101, for one additional free parameter. According to the F-test, the likelihood that this is a chance improvement is $1.06\times10^{-7}$.  This highly significant improvement strongly suggest that at these azimuths the front has a finite width.

\begin{figure}
\begin{minipage}{0.48\textwidth}
\includegraphics[width=1.4\textwidth,clip=t,angle=0.]{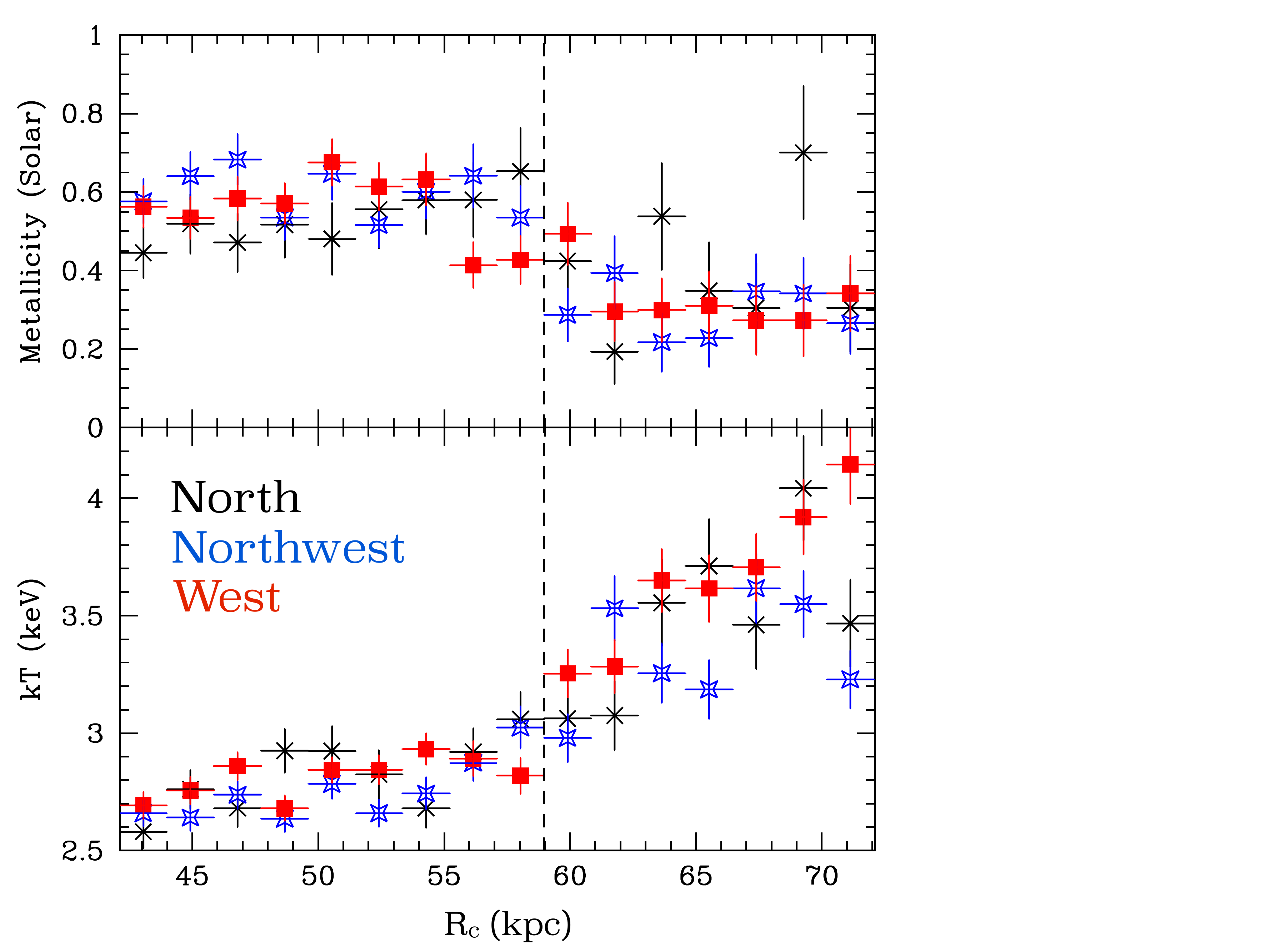}
\end{minipage}
\vspace{0.0cm}
\caption{Projected temperature and metallicity profiles along three 22.5 degree wide wedges, extracted from the northern direction westward. The profiles extracted from the north, northwest, west are shown in black, blue, and red, respectively. The best fit radius of the front determined by fitting the surface brightness profiles is indicated by the dashed line (the uncertainty on the front radius is $\pm2$~kpc). Radii are measured from the center of curvature of the cold front. } 
\label{NWprof}
\end{figure}

\begin{figure}
\begin{minipage}{0.49\textwidth}
\vspace{-2.0cm}
\hspace{-0.6cm}\includegraphics[width=1.55\textwidth,clip=t,angle=0.]{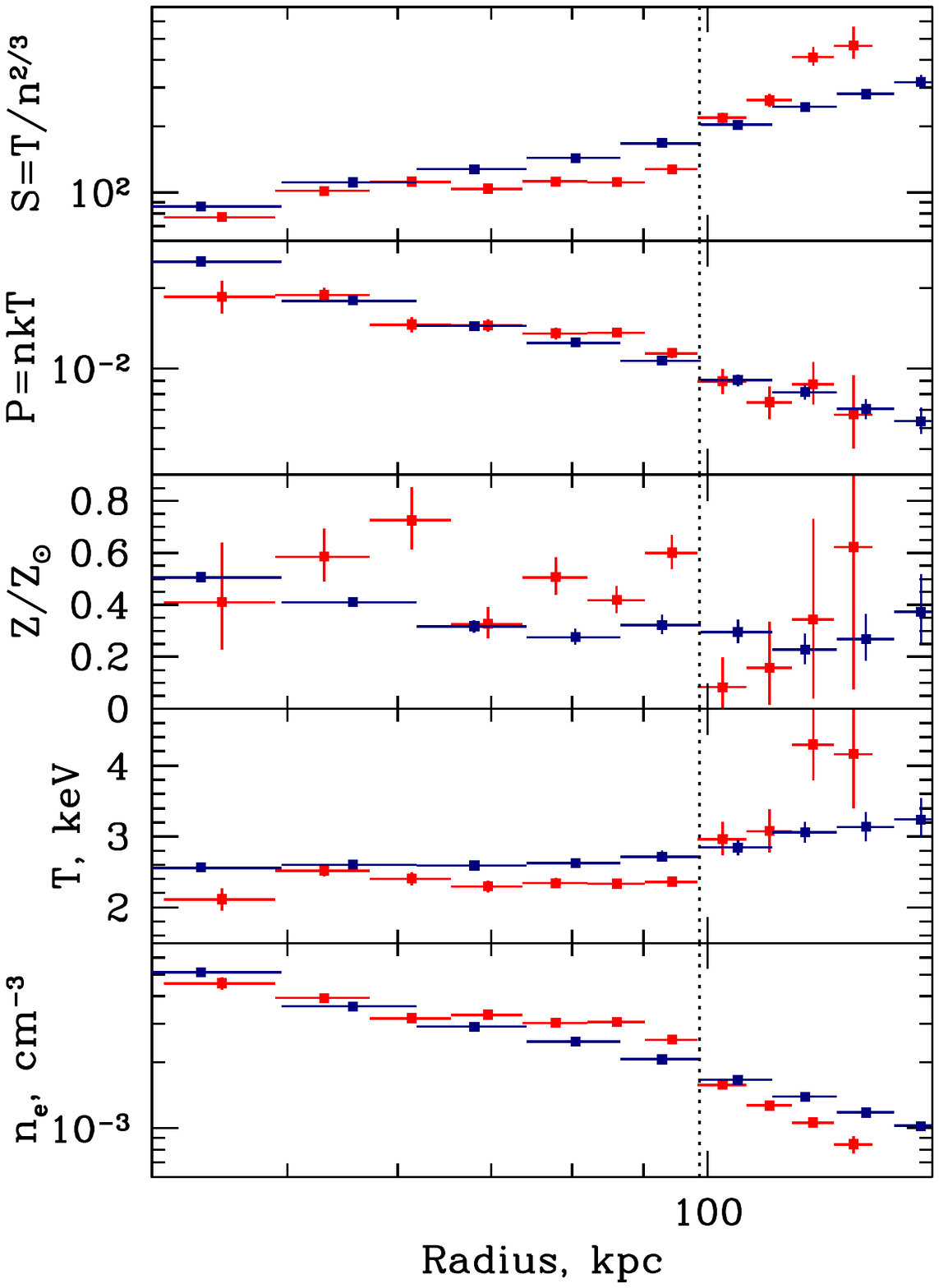}
\end{minipage}
\vspace{-3.5cm}
\caption{Deprojected electron density, temperature, metallicity, pressure, and entropy profiles (from the bottom to the top), respectively. The red datapoints were determined in a 30 degree wedge extending from the north to the west, centered on M~87, using {\it Chandra} data and the blue datapoints are averaged over the other directions (330 degree region that excludes the wedge) using {\it XMM-Newton} data. The radius of the cold front is indicated with the dotted line.  While the deprojected pressure is continuous across the cold front, the density, temperature, entropy, and metallicity distributions show relatively strong discontinuities. Note that under the cold front ($\sim20$~kpc) the ICM entropy is smaller and above the front it is higher than in the other directions.} 
\label{deproj}
\end{figure}

To study the projected thermodynamic properties and the metallicity of the ICM across the cold front, we extracted spectra along three 22.5~degree wide wedges (labeled North, Northwest, and West), the vertex of which is at the same center of curvature of the cold front as for the surface brightness profiles. The best fit projected temperature and metallicity profiles along the three wedges, determined in narrow 2~kpc wide regions, are shown in Fig.~\ref{NWprof}. The projected temperature as a function of radius increases from $\sim2.7$~keV to $\sim3.6$~keV across the interface, and the metallicity decreases from $\sim0.6$~Solar to $\sim0.3$~Solar. The projected temperature distribution appears to be smeared across the cold front. The observed gradient of the projected temperature is between $\sim4$~kpc (West) and 8~kpc (North) wide. 

\begin{figure*}
\vspace{-1.2cm}
\begin{minipage}{0.48\textwidth}
\includegraphics[width=1.12\textwidth,clip=t,angle=0.]{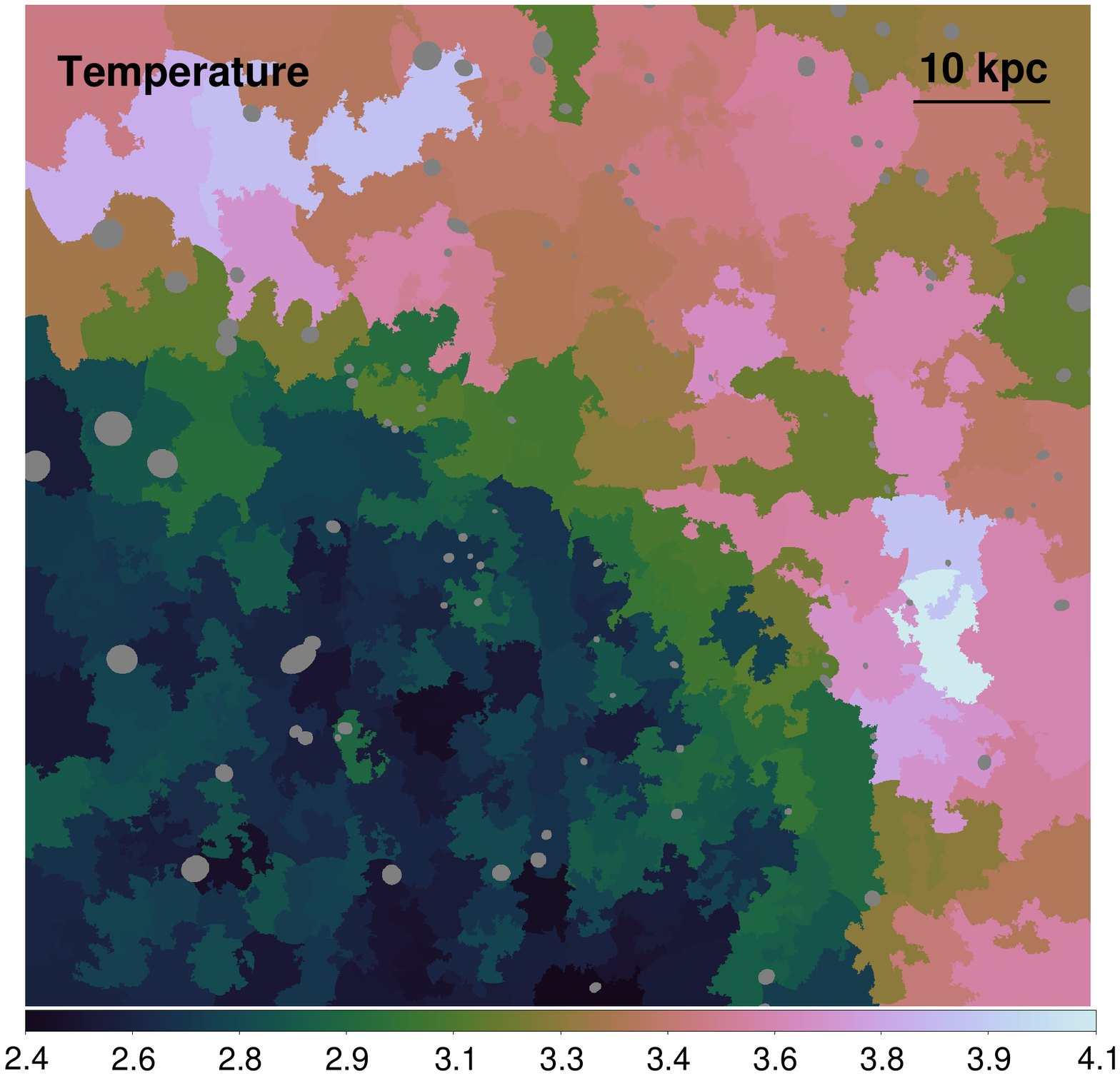}
\end{minipage}
\begin{minipage}{0.48\textwidth}
\includegraphics[width=1.12\textwidth,clip=t,angle=0.]{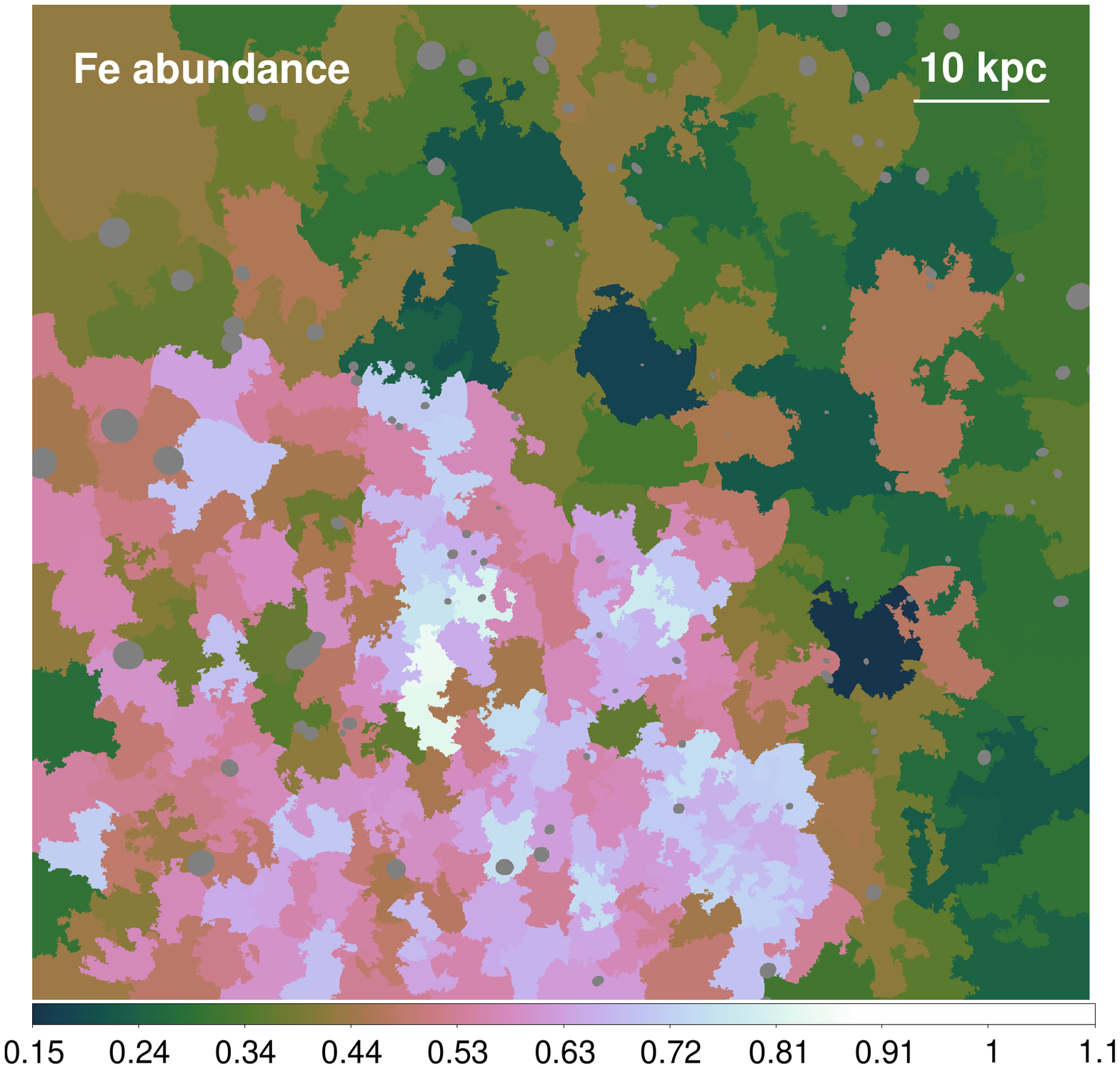}
\end{minipage}
\begin{minipage}{0.48\textwidth}
\vspace{-3.5cm}
\includegraphics[width=1.12\textwidth,clip=t,angle=0.]{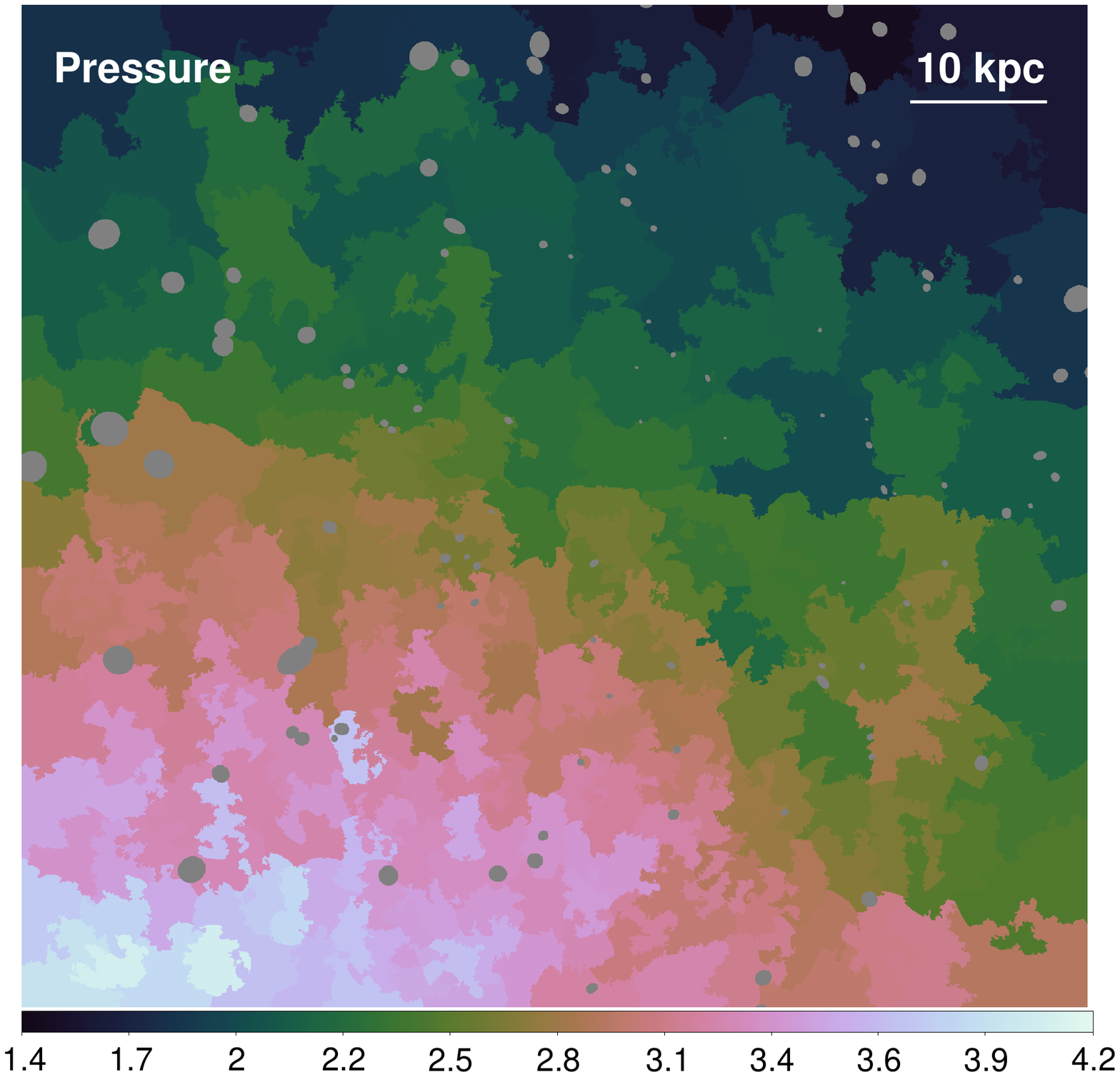}
\end{minipage}
\begin{minipage}{0.48\textwidth}
\vspace{-3.5cm}
\begin{overpic}[width=1.12\textwidth,clip=t,angle=0.]{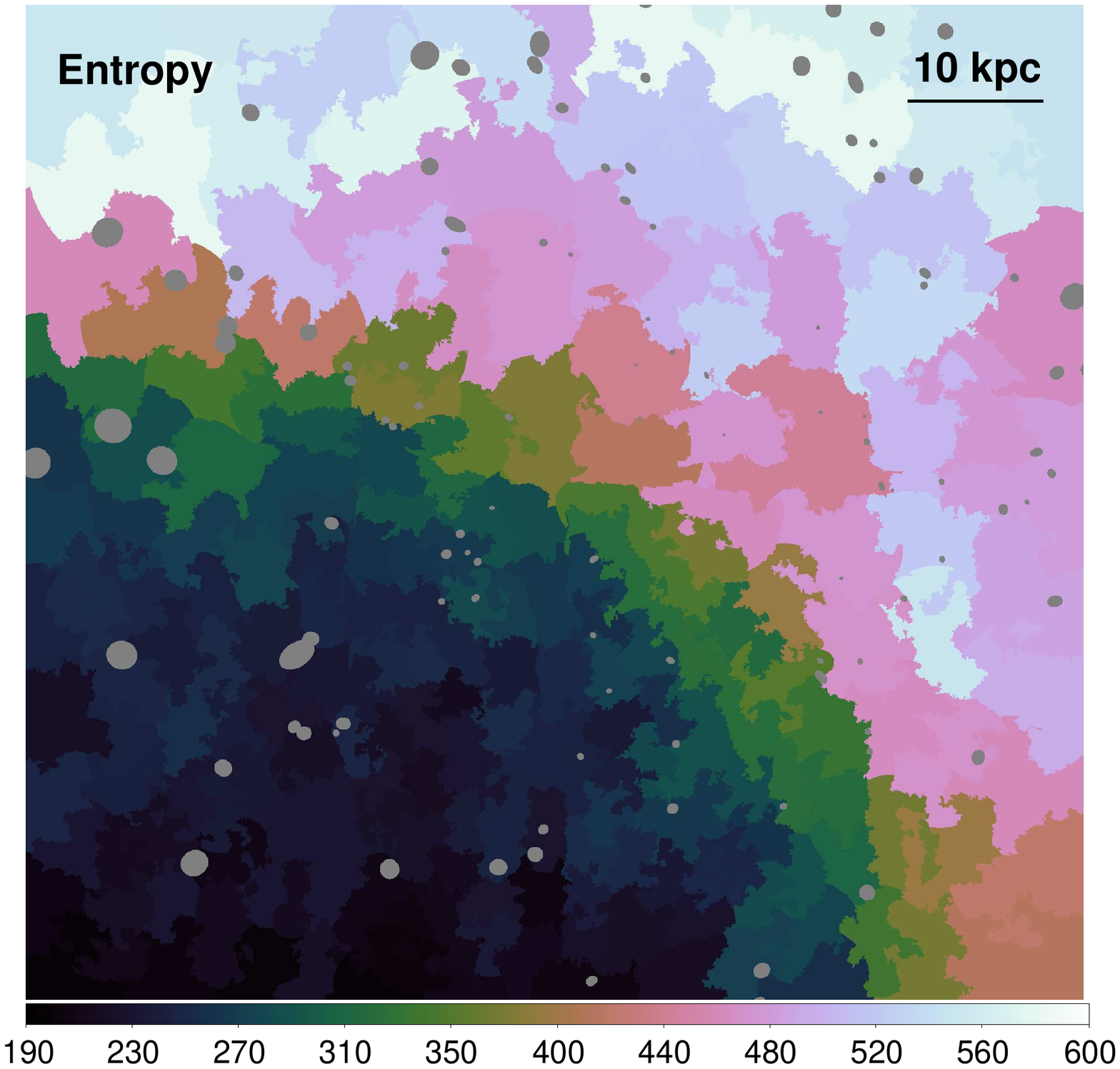}
\end{overpic}
\end{minipage}
\vspace{-1.5cm}
\caption{Projected thermodynamic maps. Each spatial region contains 5600 net counts (S/N of 75). To compare the pseudo-pressure and pseudo-entropy distributions, we assume the gas uniformly distributed along the line-of-sight depth of $l = 1$ Mpc over the entire field of view. The units of temperature (top left), Fe abundance (top right), pressure (bottom left) and entropy (bottom right) are keV, Solar \citep{lodders2003}, 1000~keV~cm$^{-3}(l/\rm{1~Mpc})^{1/2}$ and keV~cm$^2(l/\rm{1~Mpc})^{1/3}$, respectively. The fractional $1\sigma$ statistical errors are $\sim 4$ per~cent for temperature, 5 per cent for pressure and entropy, and $\sim10$--20 per cent for metallicity. Point sources were excluded from the regions shown in grey. } 
\label{maps}
\end{figure*}

The 2D maps of the projected thermodynamic properties shown in Fig.~\ref{maps} confirm these trends. While the temperature clearly increases across the front, the best fit metallicity drops from $\sim0.6$ Solar to $\sim0.3$ Solar. The projected temperature and entropy distributions appear slightly smeared at the edge. The smearing is likely to be at least in part due to projection effects (see Sect.~\ref{diffusedisc} for more discussion). As expected, the cold front is not visible in the pressure map, indicating that the sloshing gas is in approximate hydrostatic equilibrium within the cluster potential. The linear structures identified in the images (see Fig.~\ref{chandraim}) do not show any significant difference in temperature or metallicity from the surrounding ambient medium. 

In order to study the temperature difference of the quasi-linear features and the ambient plasma in more detail, we extracted five spectra from rectangular regions: three spectra from regions centered on the quasi-linear features, and two spectra from the surrounding regions. Based on fitting the combined spectra, the best fit temperature of the quasi-linear features is $kT=2.77\pm0.03$~keV and the best fit temperature of the surrounding gas is $kT=2.92\pm0.03$~keV, qualitatively consistent with the trend expected from the simulations we present in Sect.~\ref{sim} (see Fig.~\ref{beta100zoom}). However, we caution that because the surface brightness excess associated with these features is only $\sim10$~per cent, the measured temperature difference could easily be due to the underlying spatial variation in the temperature of the ambient gas.

In order to determine the deprojected spectral properties of the ICM across the cold front, we also extracted spectra from a wedge centered on M~87 at the bottom of the gravitational potential of the Virgo Cluster. Centering the series of partial annuli used for spectral extraction at the bottom of the global gravitational potential enables us to perform the most robust deprojection of the stratified cluster atmosphere. The 30 degrees wide wedge used for extracting the {\it Chandra} spectra extends from the north towards the west. For this particular wedge, the width of the cold front is the least affected by the choice of the different center. 
Fig.~\ref{deproj} shows the comparison of the  deprojected electron density, temperature, metallicity, pressure, and entropy profiles determined in the wedge using \chandra\ data (red datapoints) with the properties determined in the other directions (averaged over a 330 degree region that excludes the wedge) using {\it XMM-Newton} spectra (blue datapoints). While the deprojected pressure appears continuous across the cold front, the density, temperature, entropy, and metallicity distributions show clear discontinuities. The deprojected entropy profile extracted across the cold front shows a remarkably flat distribution underneath the discontinuity that extends across a radial range of $\sim50$~kpc.  The comparison also shows that under the cold front ($\sim20$~kpc from the discontinuity) the ICM entropy is smaller and above the front it is higher than in the other directions. The pressure appears slightly higher under the cold-front than in the other directions. We verified that the deprojected profiles determined using {\it Chandra} and {\it XMM-Newton} data in the 30 degree wedge show consistent results. We caution that while our deprojection procedure assumes spherical symmetry centered on M87, the distribution of the gas uplifted by sloshing is likely to be different and therefore our inferred thermodynamic properties underneath the discontinuity may be biased.

\section{Numerical simulations}
\label{sim}

\begin{figure*}
\vspace{-1.35cm}
\includegraphics[width=0.82\textwidth]{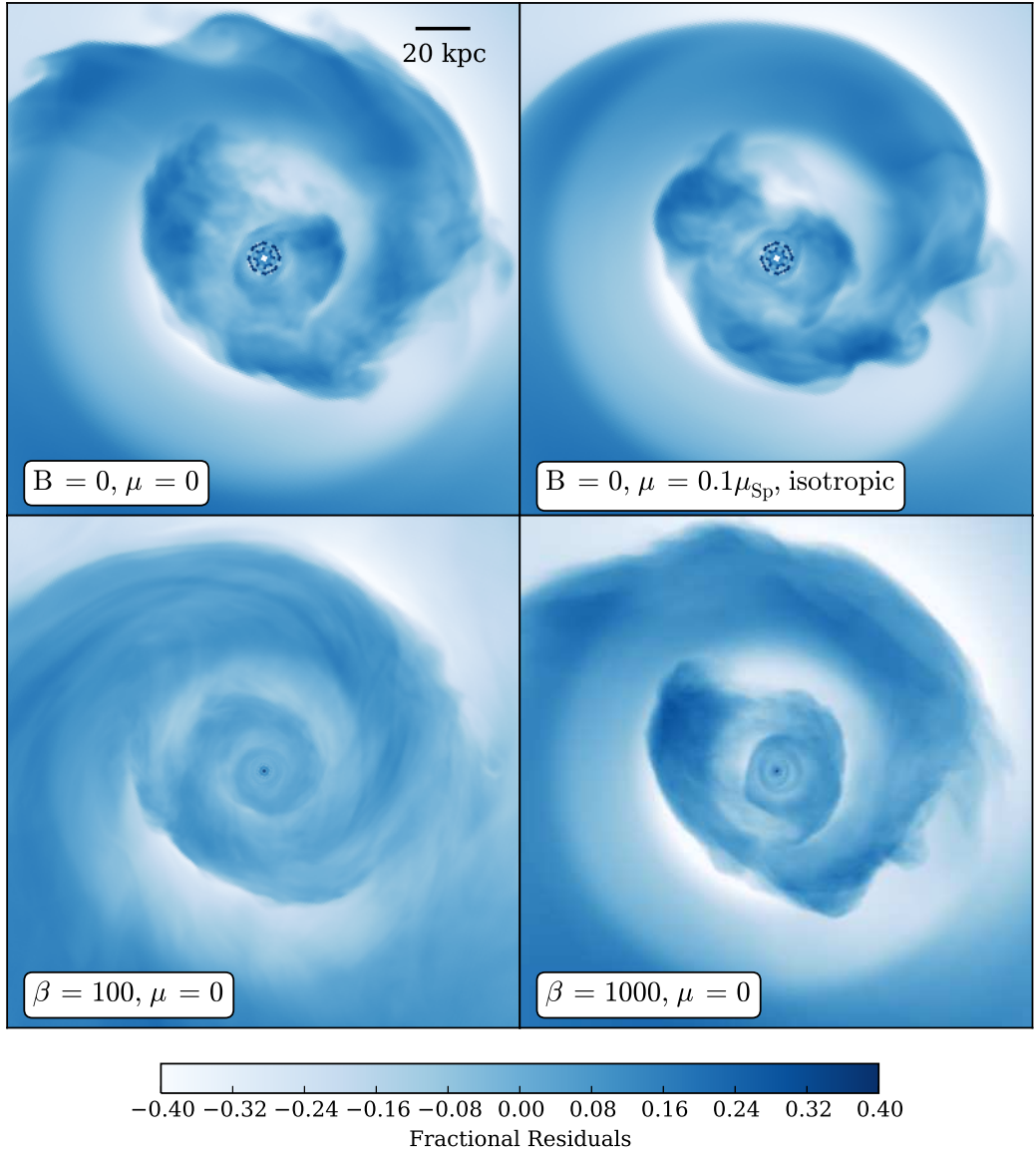}
\vspace{-1.5cm}
\caption{Simulated X-ray surface brightness residual distributions for the Virgo Cluster simulation at $t=1.7$~Gyr after the closest passage of the subcluster for two pure hydrodynamic simulation runs with different viscosities (top row), and two MHD runs with different initial plasma beta parameters (bottom row). } 
\label{4simfigs}
\end{figure*}

We compare our long exposure of the Virgo cold front to magnetohydrodynamic (MHD) simulations. 
These simulations were performed using the {\texttt{Athena 4.1}} astrophysical simulation code \citep{stone2008}. 
The technical details of the simulation setup are discussed in \citet{zuhone2015} and the initial conditions, based on the setup from \citet{rodiger2011} and \citet{rodiger2013}, were chosen to match the global Virgo Cluster potential and the ICM properties. The cluster is modeled as a cooling-core system with a mass of $\sim2\times10^{14}$~M$_{\odot}$ and $kT\sim2$~keV. The sloshing in the simulation is due to the passage of a gas-less dark-matter subcluster with a mass of $2\times10^{13}$~M$_{\odot}$ at $r\sim100$~kpc at the time of $t\sim1$~Gyr after the beginning of the simulation. We note that simulations show that similar cold front features are produced by gas-less perturbers approaching at close distances or gas-filled perturbers approaching at larger distances, indicating that the choice of perturber is not as important as the shape of the potential well and the thermodynamic profiles of the main cluster \citep{zuhone2010, rodiger2011,rodigerzuhone2012}. The time at which the properties of the Virgo cold front most resemble the results of the simulation is $\sim1.7$~Gyr after the passage of the subcluster. In Fig.~\ref{4simfigs}, we show the synthetic X-ray surface brightness residual distributions  for the simulation at this time for two pure hydrodynamic runs with different viscosities (top row), and for two MHD runs (bottom row). The pure hydrodynamic simulations were performed assuming either inviscid ICM ($\mu=0$), or 10~per cent of the isotropic Spitzer-like temperature dependent viscosity ($\mu=0.1\mu_{\rm Sp}$). Both MHD runs started with a tangled magnetic field distribution \citep[set up as in ][]{zuhone2011}, one with an initial ratio of thermal to magnetic pressure of $\beta=1000$ and the other with $\beta=100$.

The two pure hydrodynamic simulations are visibly inconsistent with the data: the inviscid run shows large, well-developed Kelvin-Helmholtz instabilities (KHI) that are not seen at the Virgo cold front  image \citep[the surface brightness profiles extracted from the mock image show a distorted front with multiple strong edges; see also Fig. 4 in ][]{rodiger2013}; and the run with 10 per cent of Spitzer viscosity produces a discontinuity that is much smoother than the profiles extracted across the northwestern part of the cold front, seen in Fig.~\ref{sbprofs}. Importantly, pure hydrodynamic simulations do not reproduce the bright linear features seen in the right panel of Fig.~\ref{chandraim}, though such features appear in the MHD run with the initial plasma $\beta=100$; we will discuss this in Section \ref{magneticdisc}.

\section{Discussion}
\label{discussion}

\begin{figure*}
\begin{center}
\vspace{-0.92cm}
\begin{minipage}{0.48\textwidth}
\hspace{-1.8cm}
\vspace{1.6cm}
\includegraphics[width=1.\textwidth,bb = -184 162 796 629]{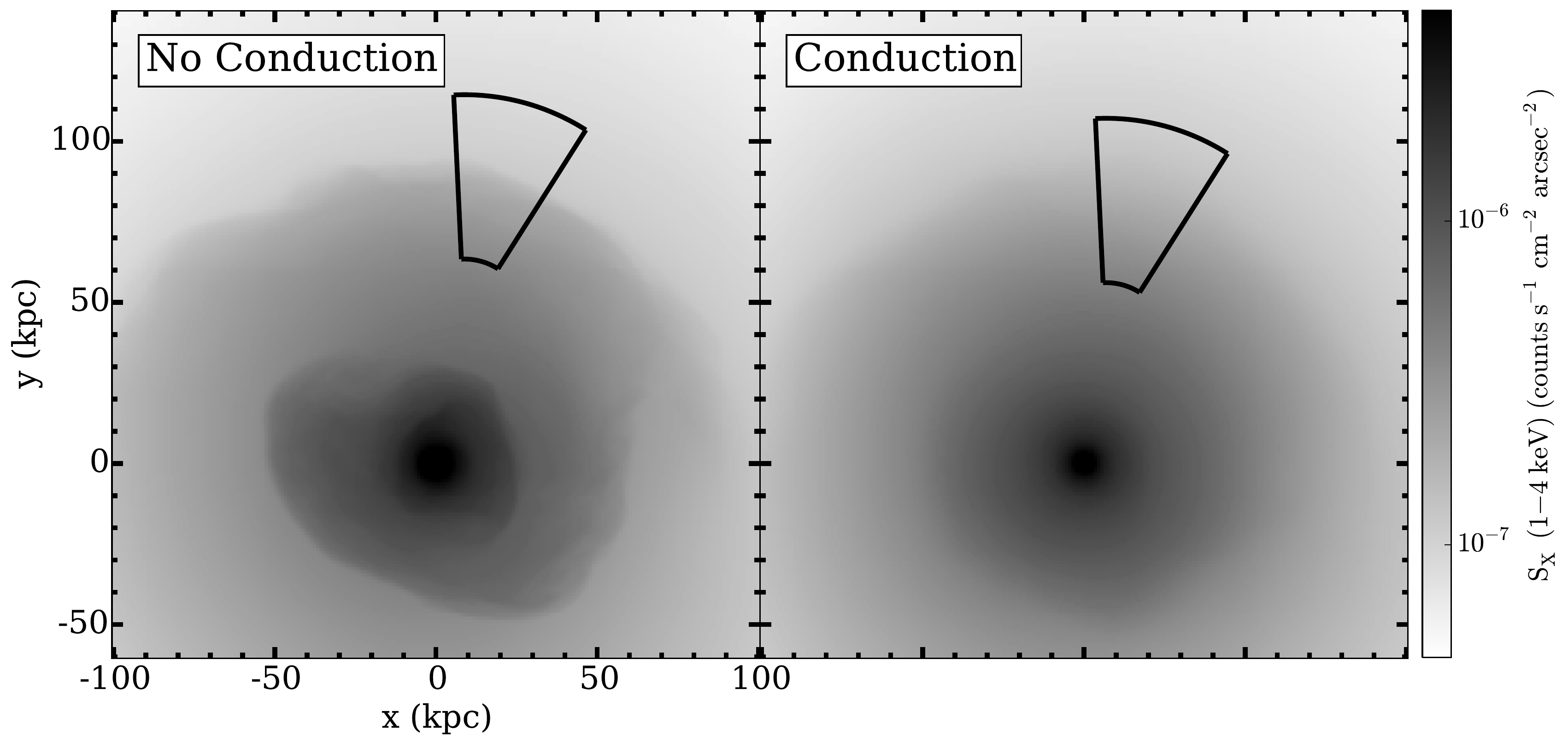}
\end{minipage}
\begin{minipage}{0.48\textwidth}
\vspace{-1.6cm}
\hspace{-1.4cm}
\includegraphics[width=0.95\textwidth, bb = -156 153 768 638]{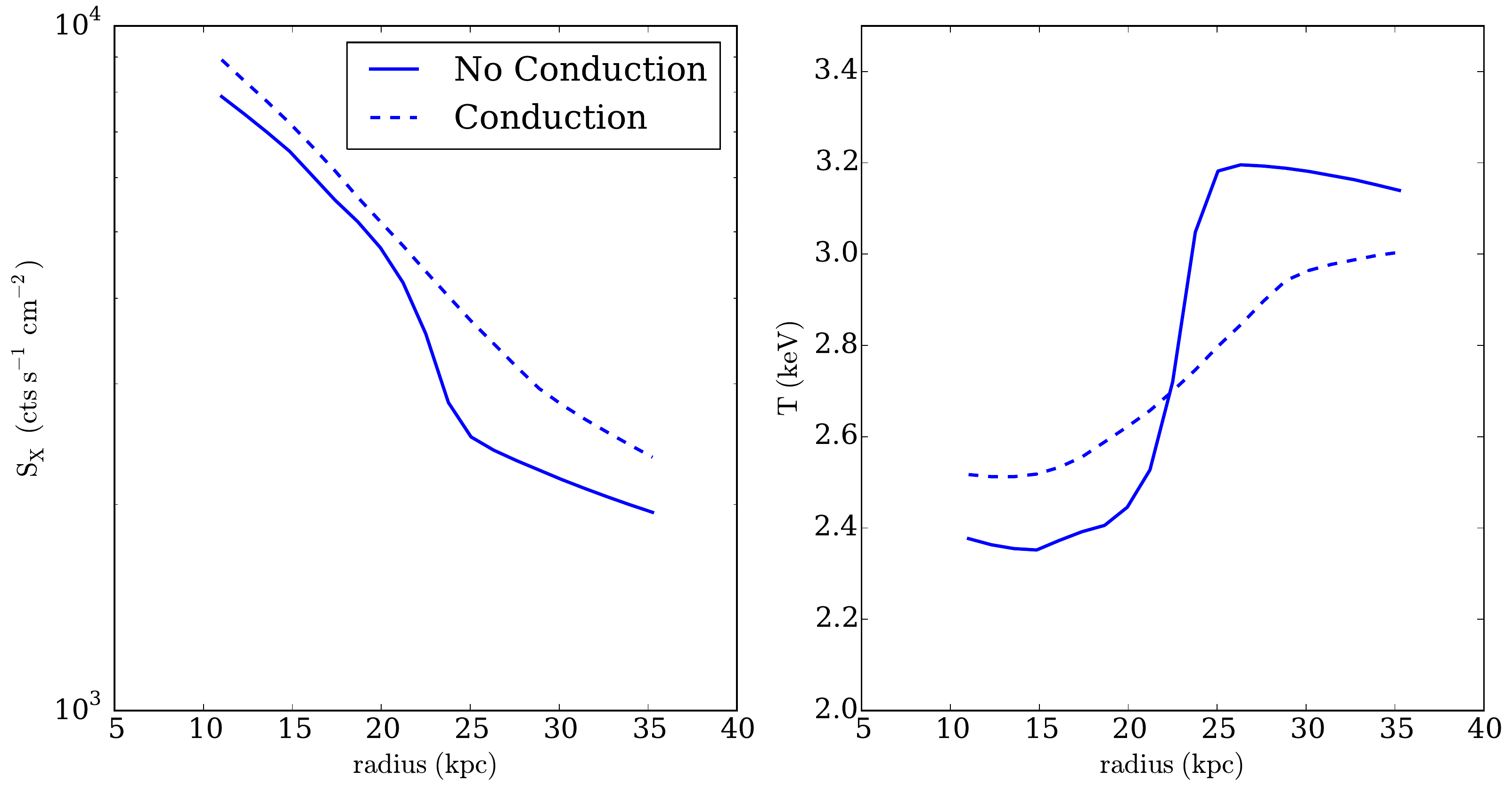}
\end{minipage}
\end{center}
\vspace{-0.4cm}
\caption{Mock surface brightness images (left two panels), and surface brightness and temperature profiles (right two panels) extracted from the black wedges for numerical MHD simulations performed assuming no conduction and full Spitzer conduction along the magnetic field lines. Spitzer conduction along the magnetic field lines would wash out the temperature discontinuity to a degree that is inconsistent with the observations. The observed temperature and surface brightness profiles (see Fig.~\ref{sbprofs}, \ref{NWprof}) are consistent with the conduction along the magnetic field lines being suppressed by a factor $\gtrsim10$.}
\label{cond}
\end{figure*}

\subsection{The width of the cold front and suppressed diffusion
\label{diffusedisc}}

\begin{figure*}
\begin{center}
\begin{minipage}{0.48\textwidth}
\includegraphics[width=1\textwidth]{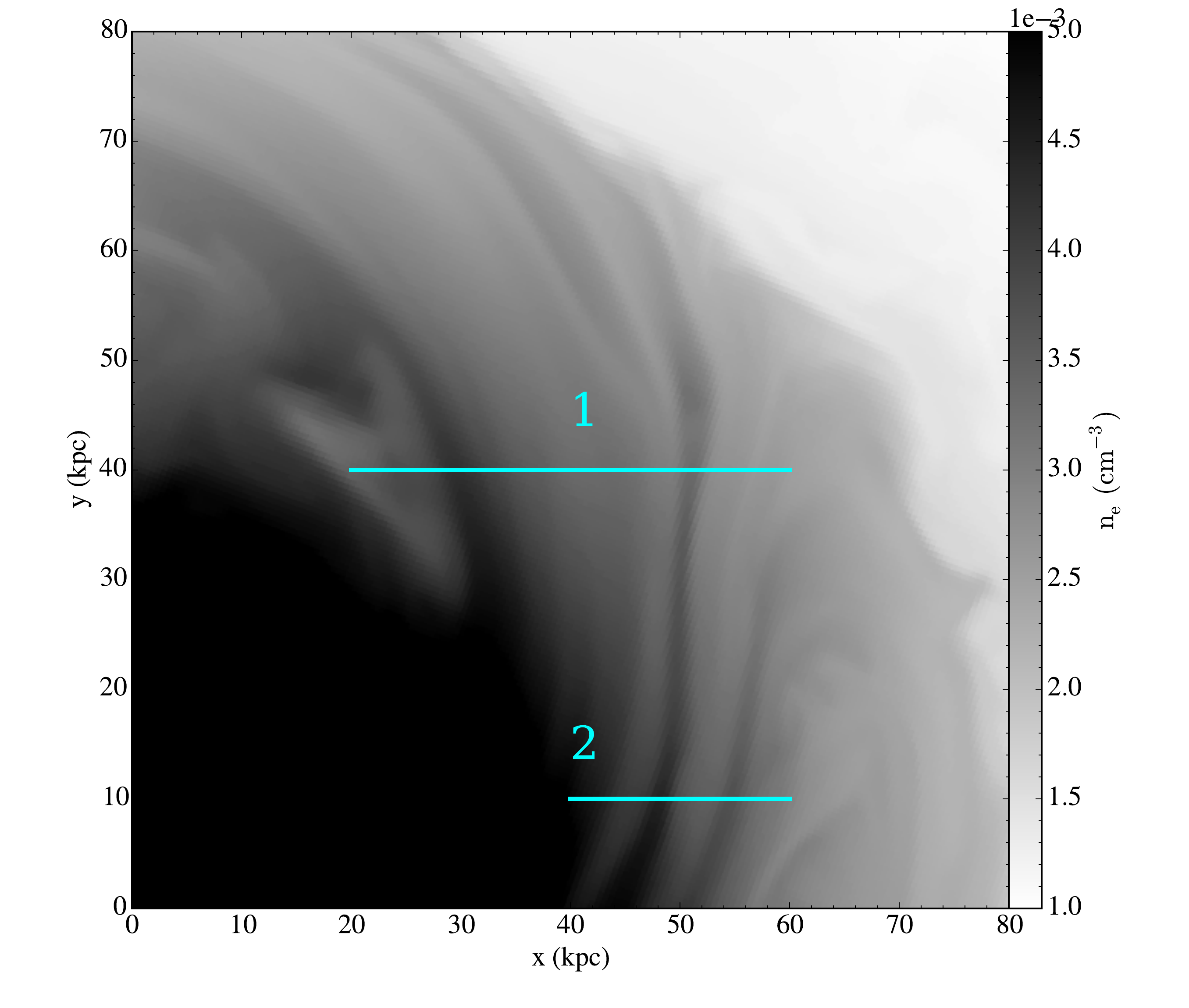}
\end{minipage}
\begin{minipage}{0.48\textwidth}
\includegraphics[width=1\textwidth]{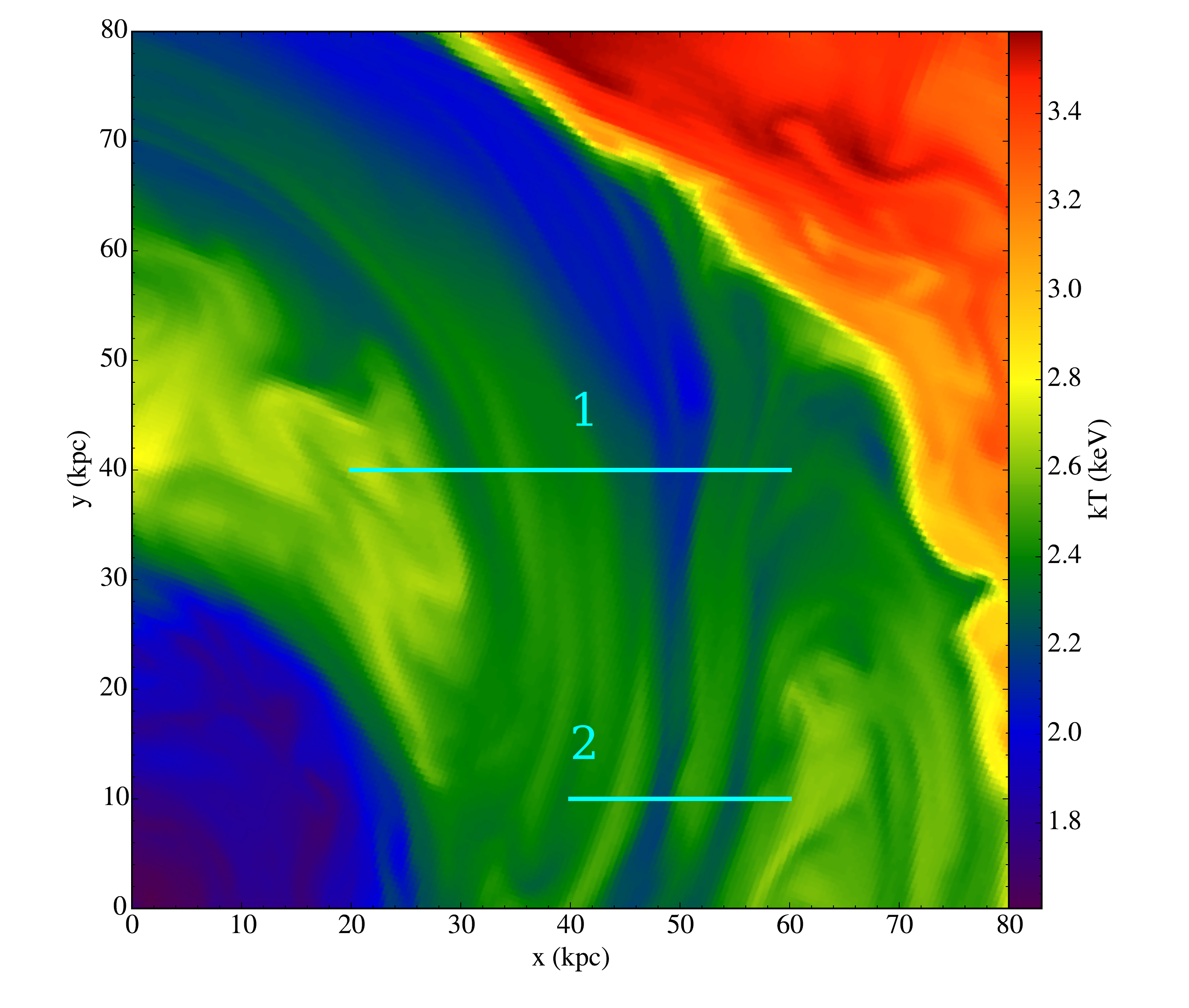}
\end{minipage}
\begin{minipage}{0.48\textwidth}
\includegraphics[width=1\textwidth]{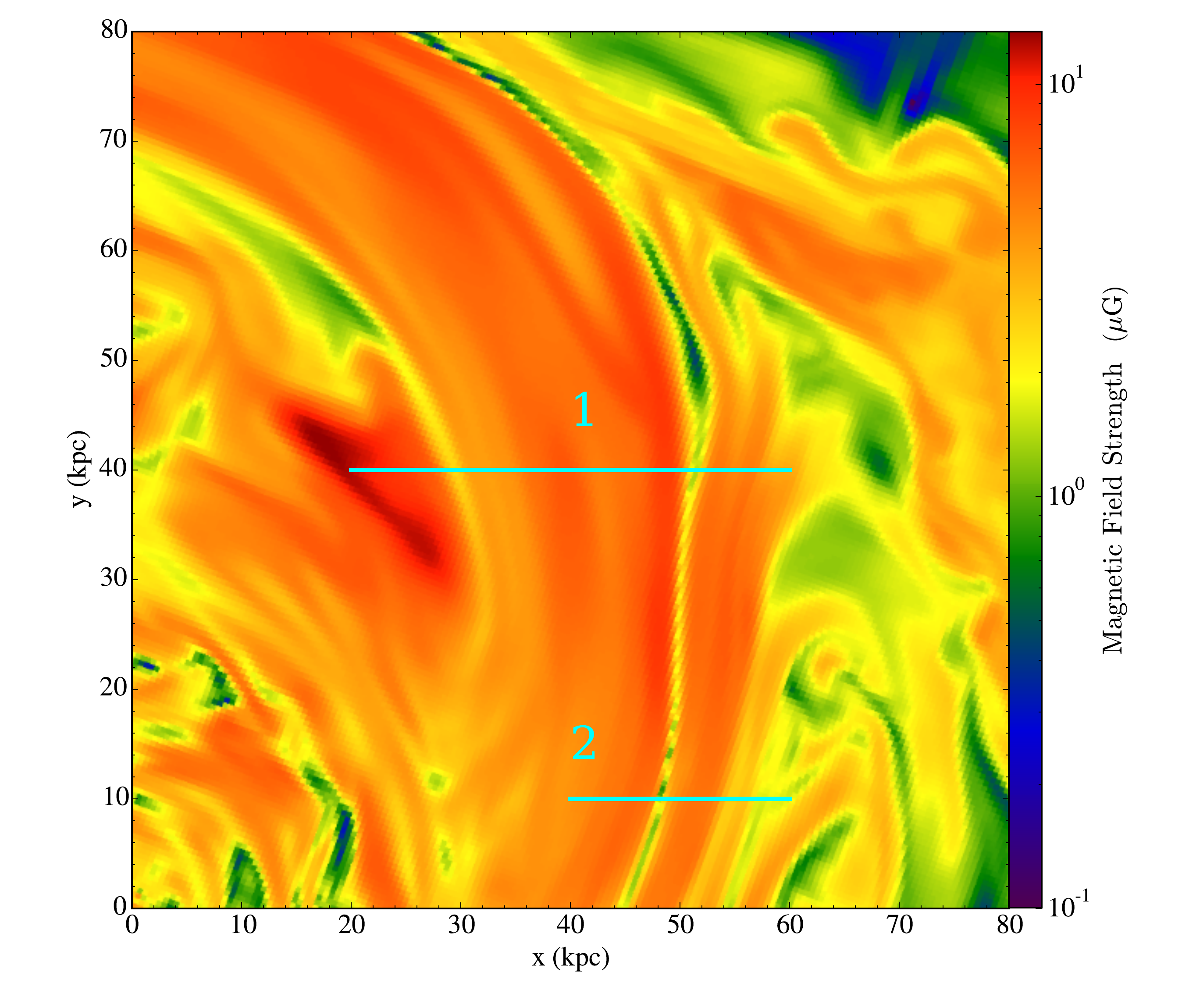}
\end{minipage}
\begin{minipage}{0.48\textwidth}
\includegraphics[width=1\textwidth]{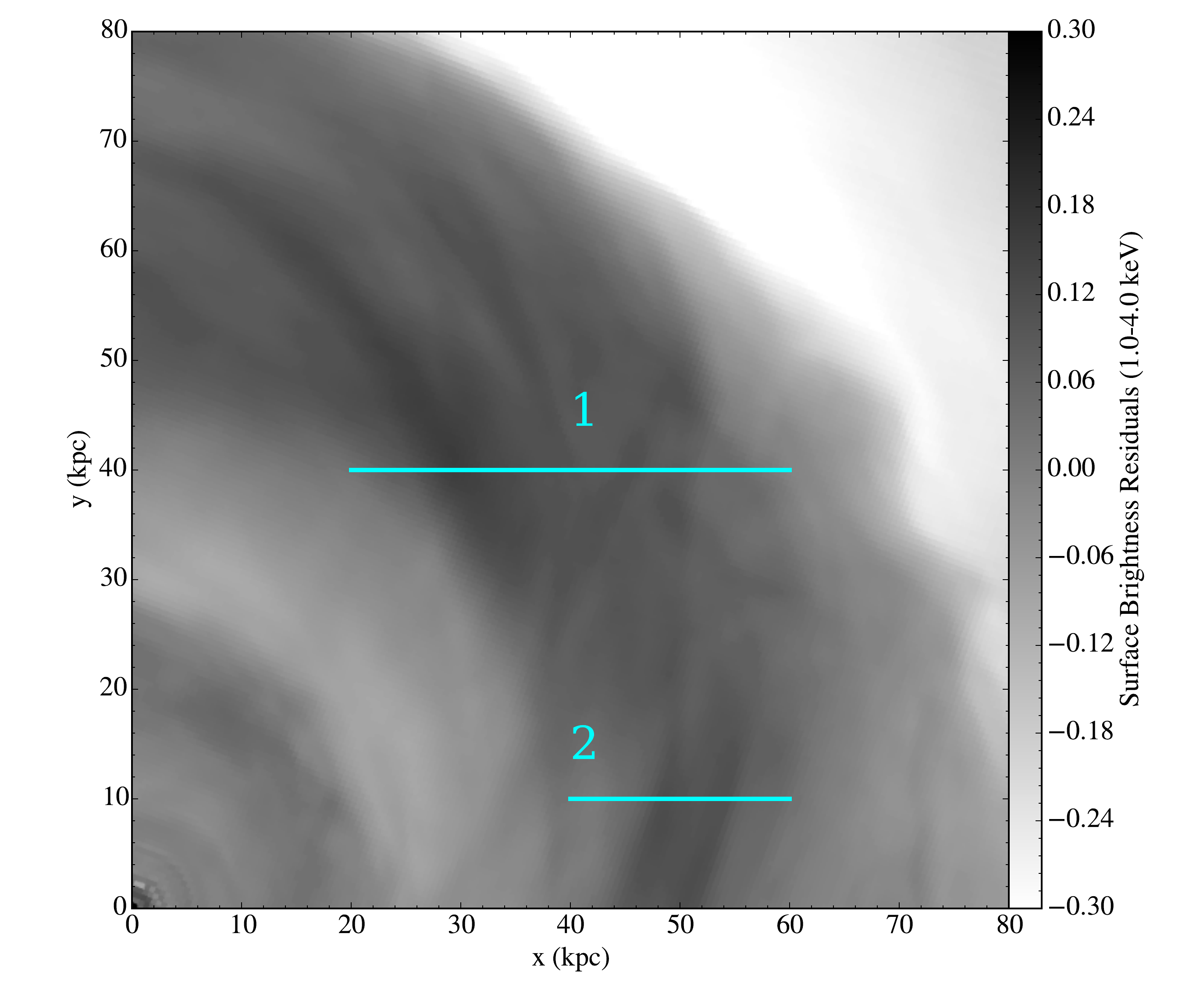}
\end{minipage}
\caption{A zoom in on a $80\times80$~kpc$^2$ region around the cold front located northwest of the cluster center for the MHD simulation run with an initial magnetic to thermal pressure ratio $\beta=100$ (see the bottom right panel of Fig.~\ref{4simfigs}). {\it Top-left:} slice of gas density; {\it top-right:} slice of gas temperature; {\it bottom-left:} slice of magnetic field strength; {\it bottom-right:} projected surface brightness residuals. The solid cyan lines indicate the extraction regions for the profiles shown in Fig.~\ref{cut2}.}
\label{beta100zoom}
\end{center}
\end{figure*}

\begin{figure*}
\begin{center}
\vspace{1.5cm}
\begin{minipage}{0.48\textwidth}
\hspace{-1.8cm}\includegraphics[width=0.95\textwidth, bb = -338 -210 950 1002]{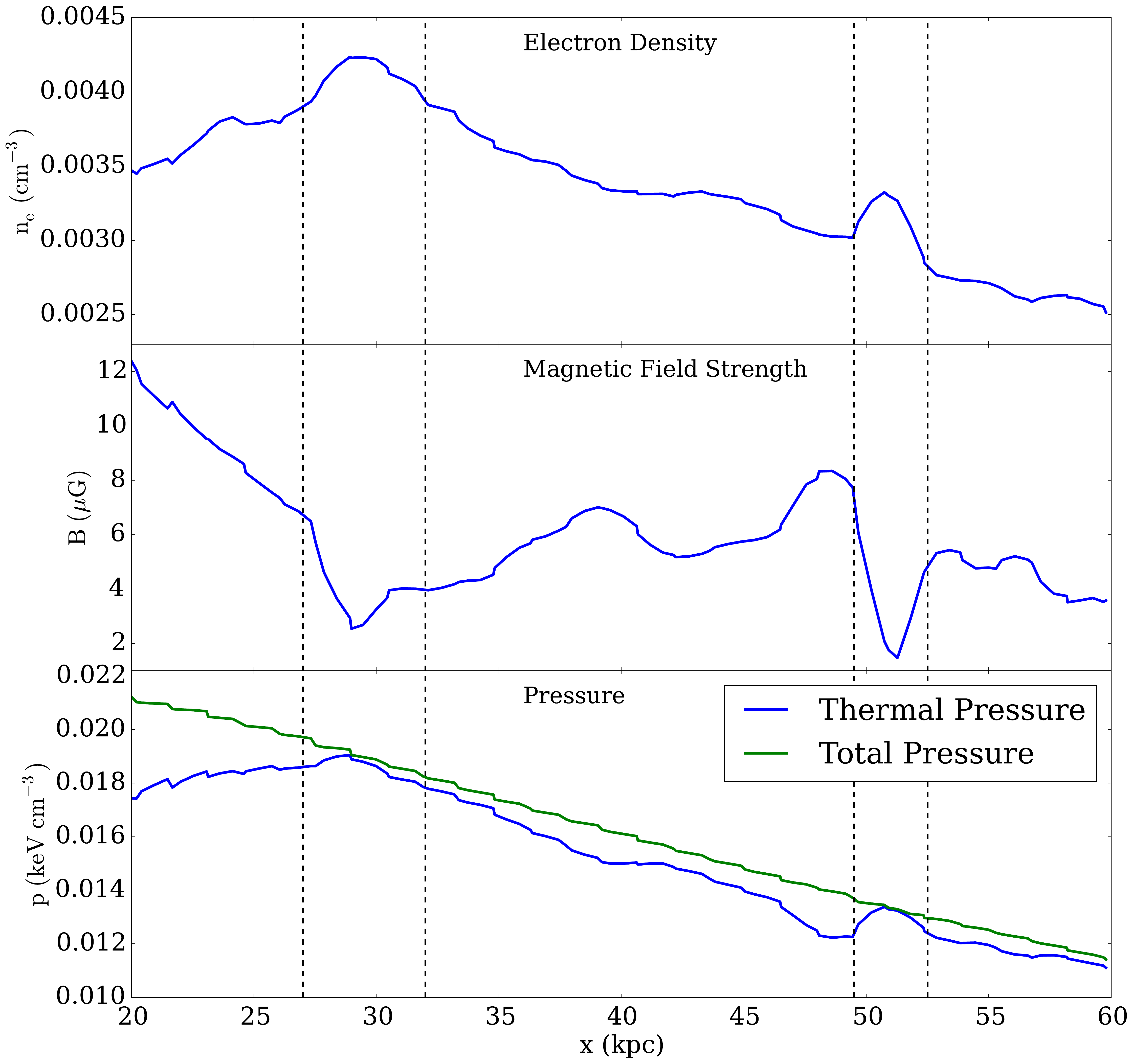}
\end{minipage}
\begin{minipage}{0.48\textwidth}
\hspace{-1.8cm}\includegraphics[width=0.95\textwidth, bb = -338 -210 950 1002]{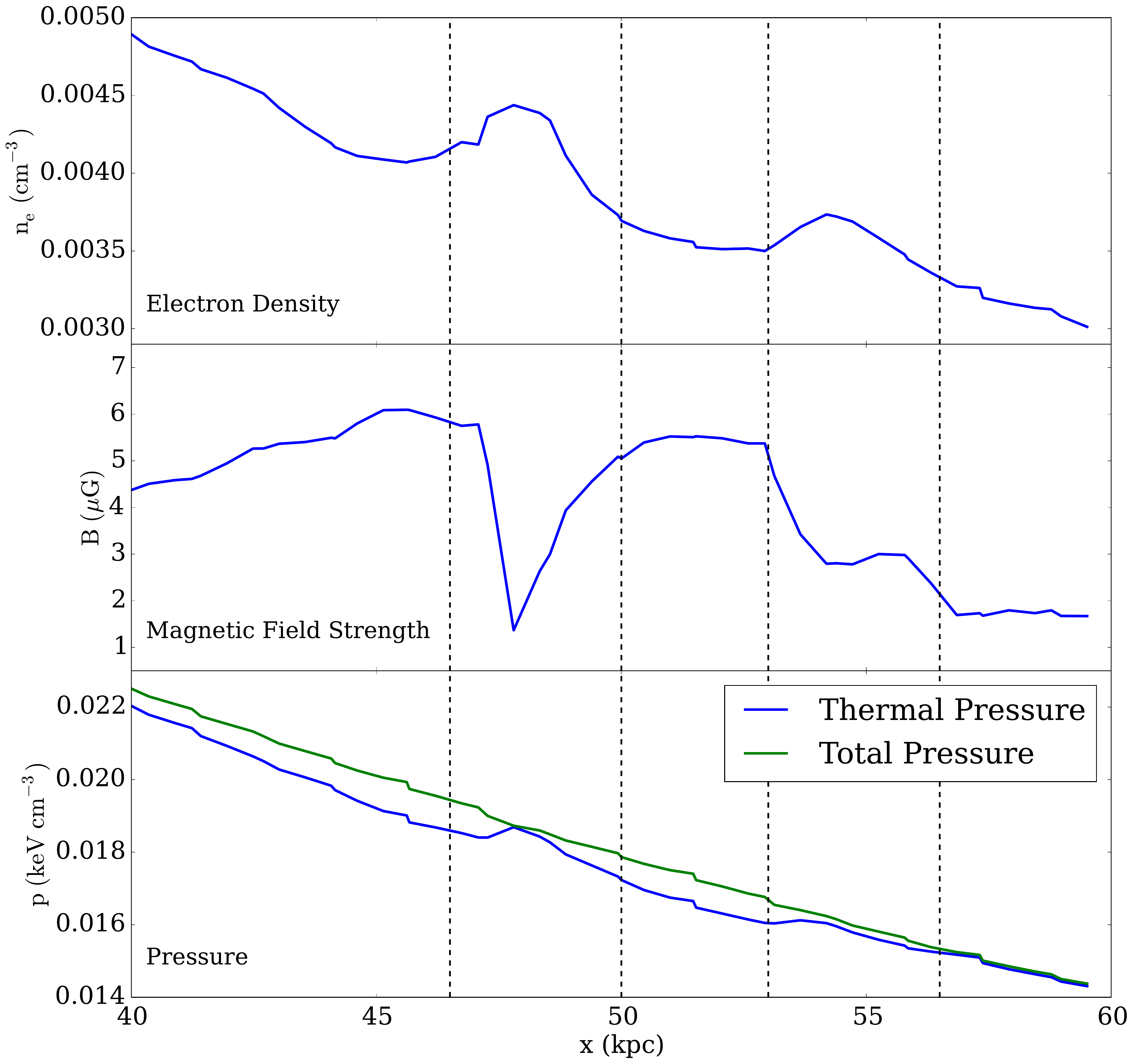}
\end{minipage}
\vspace{-1cm}
\caption{Profiles of electron density, magnetic field strength, and pressure along the cyan lines indicated in Fig.~\ref{beta100zoom} as `1' (left panel) and `2' (right panel). The difference between the total and thermal pressure is due to magnetic pressure support. Dashed lines indicate approximate boundaries of density enhancements along the profile.}
\label{cut2}
\end{center}
\end{figure*}

The observed surface brightness profiles across the cold front show considerable azimuthal variation. While at certain azimuth angles (particularly in the north) the profiles are relatively well described by the model of a spherically symmetric cloud with a power-law density distribution, in other directions they deviate from the model, are wider, and the best fit front parameters show significant azimuthal variation. 

For the best fit deprojected temperatures and electron densities, the Coulomb mean free paths inside and outside of the cold front are $\lambda_{\rm in}=0.77$~kpc and $\lambda_{\rm out}=2.2$~kpc, respectively. For unsuppressed diffusion, particles diffusing from the bright, dense side of the discontinuity to the low density outer ICM will be especially efficient at broadening the cold front. The Coulomb mean free path for diffusion from the inner bright side to the outer faint side of the discontinuity is:
 \begin{equation}
 \lambda_{\rm in \rightarrow out}=15\left(\frac{T_{\rm out}}{\rm 7~keV}\right)^2\left(\frac{n_{\rm e, out}}{10^{-3}~\rm{cm}^{-3}}\right)\left(\frac{T_{\rm in}}{T_{\rm out}}\right)\frac{G(1)}{G(\sqrt{T_{\rm in}/T_{\rm out}})}=1.7~\rm kpc,
 \end{equation} 
where $n_{\rm e, out}$ is the electron density outside the cold front, $T_{\rm in}$ and $T_{\rm out}$ are the temperatures inside and outside the cold front, and $G(x)=[\phi(x)-x\phi'(x)]/2x^2$, where $\phi(x)$ is the error function \citep{spitzer1962,markevitch2007}. 
Our conservative upper limit of 2.5~kpc on the intrinsic width of the cold front, determined in the sector with the sharpest brightness edge, corresponds to about 1.5 Coulomb mean free paths.

Part of the apparent broadening of the cold front in the Virgo Cluster may be caused by projection effects and the true intrinsic width of the discontinuity is also likely to be smaller than the Coulomb mean free path. Because the front is seen along the surface in projection, any deviations from the ideal spherical shape will smear the edge. Therefore, the observed width is a conservative upper limit on the actual width of the discontinuity. Based on comparisons to numerical simulations, \citet{rodiger2011,rodiger2013} argue for sloshing in the Virgo Cluster close to the plane of the sky, ruling out orbital planes that are further than 45 degrees from the plane of the sky.  We verified that the mock X-ray images of sloshing in our simulations, produced for various lines of sight, are consistent with those in \citet{rodiger2011,rodiger2013}. For sloshing taking place predominantly in the plane of the sky projection effects are minimized. Their smearing effect gets more important as the shape of the front departs from spherical symmetry and as the inclination of the orbital plane to the plane of the sky gets larger (until the inclination approaches 90 degrees, when sloshing is along the line-of-sight).

Unsuppressed diffusion would smear the density discontinuity by {\emph{several}} mean free paths. Our upper limit on its intrinsic width therefore indicates that diffusion is suppressed across the cold front. This is consistent with the results obtained for the cold front identified with a subcluster merger in Abell~3667, where \citet{vikhlinin2001} showed that the gas density discontinuity is several times narrower than the Coulomb mean free path \citep[see][for updated results]{markevitch2007}. \citet{churazovInogamov2004} argue that a small but finite width of the interface set by diffusion, that is of the order of a few per cent of the curvature radius ($\sim2$~kpc for the Virgo cold front), could strongly limit the growth of perturbations such as KHI \citep[see also][]{chandra1961}. Given projection effects that are expected to be broadening the observed width of the discontinuity, the intrinsic width of the Virgo cold front is likely to be too small to be suppressing the development of KHI.

\subsection{Kelvin-Helmholtz instabilities and ICM viscosity}

The primary effect of projection is the broadening of the surface brightness edge and possibly a relatively smooth azimuthal change of the broadening. 
However, the non-zero width of the cold front, combined with the observed substructure and the azimuthal variations of the best fit radius and density jump seen in Fig.~\ref{sbprofs}, require the presence of additional processes, such as KHI. The temperature profiles shown in Fig.~\ref{NWprof} are consistent with the presence of instabilities on a scale of a few kpc. 


Constrained hydrodynamic simulations by \citet{rodiger2012,rodiger2013} demonstrate the efficiency of viscosity at suppressing KHI and show that for viscosities of about 10 per cent of the Spitzer value or larger, at this cold front, KHIs should be suppressed. Our pure hydrodynamic simulation run with 10 per cent Spitzer viscosity is consistent with these conclusions. Based on comparison to these numerical simulations, the presence of KHI would imply that the effective viscosity of the ICM is suppressed by more than an order of magnitude with respect to the isotropic Spitzer-like temperature dependent viscosity. Our conclusions, however, depend on the correctness of the assumed merger model. 


The growth of KHIs may be facilitated by the presence of magnetic fields which lower the effective viscosity of the plasma by preventing the diffusion of momentum and heat perpendicular to the field lines. Also, micro-scale MHD plasma instabilities may set an upper limit on the viscosity that is much lower than that expected from ion collisionality \citep{kunz2014}.  However, depending on the field geometry and strength, magnetic fields may also protect cold fronts against instabilities. A layer of strong enough magnetic fields aligned with the cold front could suppress the growth of KHIs \citep{vikhlinin2002,keshet2010}. However, the suppression of KHI in the simulations of \citet{zuhone2011} is often partial, indicating that magnetic fields cannot always stabilize cold fronts.  Our example in the bottom-right panel of Figure~\ref{4simfigs} shows that if magnetic fields are initially weak ($\beta = 1000$ in our example), they will be unable to suppress KHI. Stronger magnetic fields (see bottom-left panel of Fig.~\ref{4simfigs} for an example with initial $\beta = 100$) suppresses most of the KHI but not all.

Other possible explanations for the broadening and the observed substructure in the surface brightness profiles include plasma depletion due to regions of amplified magnetic fields below the cold front, jet inflated bubbles entrained by the sloshing gas, or clumpy, inhomogeneous ICM distribution.

\subsection{Gas mixing at the cold front}

Well developed, long lasting, large KHI would mix the high metallicity ICM ($\sim0.6$ Solar) on the bright side of the interface with the low metallicity gas ($\sim0.3$~Solar) on the faint side. The relatively sharp (narrower than 6~kpc) metallicity gradient indicates that if the surface brightness substructure is due to KHI, they are on relatively small, few kpc, scales and are unable to efficiently mix metals across wider regions around the interface. Since cold fronts are transient wave phenomena, the lack of significant mixing across the front is not entirely surprising. Numerical simulations indicate that metallicity jumps remain preserved even for inviscid ICM  \citep{rodiger2011} and they get washed out only on very small scales, consistent with our observation and observations of sloshing cold fronts in other more distant clusters which are also associated with metallicity discontinuities \citep[see e.g.][]{ghizzardi2014}.

\subsection{Conduction across the discontinuity}

\citet{zuhone2013} show that while sloshing approximately aligns the magnetic field lines with the cold front surface, perfect alignment is not expected. KHI will re-tangle the field lines, restoring the heat flow between the gas phases above and below the cold front. 
\citet{zuhone2015} also performed simulations for the Virgo Cluster with anisotropic conduction, assuming full Spitzer conductivity along the magnetic field lines. In these, the temperature and surface brightness gradients are washed out to such a degree that is inconsistent with our observation (see Fig~\ref{cond}). The relatively narrow temperature gradients across the interface (see Fig.~\ref{NWprof}) suggest that the effective conductivity {\it along} the field lines must be reduced by effects not modeled by the simulation, such as strong curvature of the field lines, stochastic magnetic mirrors on small scales, and/or microscopic plasma instabilities \citep[e.g.][]{chandran1999,malyshkin2001,narayan2001,schekochihin2005,schekochihin2008}. While the observed temperature and surface brightness profiles are consistent with no conduction along the magnetic field lines (see Fig.~\ref{cond}), conductivity at or below 0.1 of the Spitzer value along the field lines \citep[which would still result in a steep gradient across the discontinuity, see][]{zuhone2013} cannot be excluded and the exact suppression factor remains unknown. Like our conclusions on viscosity, this result also depends strongly on the correctness of the assumed merger model.

\subsection{Amplified magnetic fields underneath the cold front
\label{magneticdisc}}

The {\it Chandra} images in Fig.~\ref{chandraim} show quasi-linear features that are $\sim10$ per cent brighter than the surrounding gas.
The simulations of \citet{zuhone2011} and \citet{zuhone2015} show that the magnetic field layers amplified by sloshing may be wide and extend relatively far below cold fronts. This can be seen clearly in the Virgo simulation with the initial plasma $\beta=100$ in Fig.~\ref{beta100zoom}, which shows slices through a $80\times80$~kpc$^2$ region around the cold front for gas density, temperature, and magnetic field strength, as well as residuals of projected surface brightness. Along the lines indicated as ``1" and ``2", we calculate the distributions of the density, temperature, magnetic field strength and pressure and show them in Fig.~\ref{cut2}. The quasi-linear features seen in the simulation are very similar to those in our observation, both in terms of width and apparent surface brightness excess. We find that regions of high density/surface brightness correspond to regions of relatively low magnetic field in the map. The thin regions of high density are surrounded by wider bands which are highly magnetized. Here, the increased magnetic pressure has pushed gas out, decreasing its density. This opens up the intriguing possibility that the observed bright bands may be due to denser ICM layers between bands of more magnetized plasma, where the magnetic pressure is amplified to $\sim5$--10 per cent of the thermal pressure. 
Though the magnetic fields in this simulation are strong, they do not completely prevent the development of KHI, consistent with the indications of KHI in our observation. 

By extrapolating the pressure profiles outside of cold fronts inwards, up to the edge, \citet{reiss2014} infer that in the inner parts of the spirals most sloshing cold fronts show a pressure jump of $P_{\rm in}/P_{\rm out}\sim 0.8$, which may correspond to the amplification of magnetic fields mainly below the discontinuity. The Virgo cold front does not show any significant pressure jump, which may be due to the amplification of the magnetic fields both above and below the front. Fig.~\ref{beta100zoom} shows that such a scenario does occur in our simulation. The magnetic fields above the cold front get amplified due to the shear in the large-scale velocity field produced by the encounter with the subcluster. The same large scale gas motions then draw the amplified magnetic fields inward, towards the cold front. 

The observed quasi-linear features may potentially also be due to compressed thermal gas formed in front of bubbles entrained by the magnetized sloshing gas, spiraling outwards below the cold front. Such bubble entrainment was previously suggested by \citet{keshet2012}. Future numerical simulations, that include both jet inflated bubbles and sloshing, will test this possible interpretation. 

Alternatively, for geometries where the sloshing occurs along our line-of-sight, KHI would develop predominantly in the plane of the sky, forming rolls projected on the bright side of the front. For such geometries, Kelvin-Helmholtz rolls projected on the sloshing surface could, in principle, appear as quasi-linear features. However, based on the comparison of the numerical simulations to the observations, such geometry is unlikely. None of the simulations that we performed show KHI in projection that would look like the observed quasi-linear features.

\subsection{The flow of the sloshing gas}

Under the cold front interface, the ICM entropy is lower and above the interface it is higher than the radial average. The low entropy gas under the interface also has roughly two times higher metallicity than the gas above the interface.  This entropy and metallicity distribution indicates convergent gas flows at the cold front interface, with the low entropy metal rich gas moving outward and, in the same time, the high entropy gas moving inward. Such convergent flows are also seen in numerical simulations \citep[e.g.][]{ascasibar2006,rodiger2011}. 

The entropy profile shows a remarkably flat distribution underneath the cold front, extending across a radial range of $\sim50$~kpc. The flat entropy distribution may indicate that, in this region, the gas uplifted by sloshing originated at the same radius. Similar entropy plateaus have also been seen in e.g. the Perseus and Abell 2142 clusters and interpreted as evidence for gas uplift by sloshing \citep[][]{simionescu2012,rossetti2013}. However, if the distribution of the uplifted gas deviates significantly from spherical symmetry centered on M87 (as assumed by our deprojection analysis), then the shape of the entropy profile measured underneath the cold front may be biased.

\section{Conclusions}
\label{conclusions}
We have analyzed a new, very deep (500~ks) \chandra\ observation, together with complementary archival {\it XMM-Newton} data, and performed tailored numerical simulations to study the nearest cluster cold front in the sky. We find that:

\begin{itemize}

\item The northern part of the front appears the sharpest, with a width smaller than 2.5~kpc (1.5~Coulomb mean free paths; at 99 per cent confidence). Everywhere along the front, the temperature discontinuity is narrower than 4--8~kpc and the metallicity gradient is narrower than 6~kpc, indicating that diffusion, conduction and mixing are suppressed across the interface. Such transport processes can be naturally suppressed by magnetic fields aligned with the cold front. We also find that the temperature jump across the cold front is inconsistent with the temperature jumps produced in simulations with anisotropic thermal conduction, indicating that conduction is also suppressed along the magnetic field lines.

\item However, the northwestern part of the cold front is observed to have a nonzero width. The broadening is consistent with the presence of projected KHI eddies on length scales of a few kpc. Based on comparison with simulations, the presence of KHI would imply that the effective viscosity of the ICM is suppressed by more than an order of magnitude with respect to the isotropic Spitzer-like temperature dependent viscosity.  

\item Underneath the cold front, we observe linear features that are $\sim10$ per cent brighter than the surrounding gas and are separated by $\sim15$ kpc from each other in projection. Comparison to tailored numerical simulations suggests that the observed phenomena may be due to the amplification of magnetic fields by gas sloshing in wide layers below the cold front, where the magnetic pressure reaches $\sim5$--10 per cent of the thermal pressure, reducing the gas density between the bright features.

\item The deprojected entropy distribution and the metallicity of the gas indicate convergent gas flows at the cold front interface, with the low entropy metal rich gas moving outward and, in the same time, the high entropy gas moving inward.

\end{itemize}

\section*{Acknowledgments}
NW thanks G. Ogrean, P. Nulsen, O. Urban, and R. Canning for discussions. Support for this work was provided by the National Aeronautics and Space Administration through Chandra Award Number GO3-14142A issued by the Chandra X-ray Observatory Center, which is operated by the Smithsonian Astrophysical Observatory for and on behalf of the National Aeronautics Space Administration under contract NAS8-03060. JAZ acknowledges support from NASA though subcontract SV2-8203 to MIT from the Smithsonian Astrophysical Observatory. Analysis of the simulation data was carried out using \texttt{yt}, a visualization and analysis software suite for simulations in astrophysics \citep[http://yt-project.org,][]{turk2011}. YI is financially supported by a Grant-in-Aid for Japan Society for the Promotion of Science (JSPS) Fellows. E. R. acknowledges the support of the Priority Programme Physics of the ISM of the DFG (German Research Foundation) and a visiting scientist fellowship of the Smithsonian Astrophysical Observatory. U.K. is supported by the European Union Seventh Framework Programme, by an IAEC-UPBC joint research foundation grant, and by an ISF-UGC grant. This work was supported in part by the U.S. Department of Energy under contract number DE-AC02-76SF00515. 

\bibliographystyle{mnras}
\bibliography{clusters}

\end{document}